\newcommand\msun{{\,M_\odot}}

\newcommand\zsun{{\rm \,Z_\odot}}

\newcommand{\unit}[1]{\ensuremath{\, \mathrm{#1}}}

\newcommand{\Ang}{~\mbox{\AA}}

\newcommand{\cMpc}{~\mbox{comoving}~\mbox{Mpc}}

\newcommand{\cmci}{~\mbox{cm}^{-3}}

\newcommand{\K}{~\mbox{K}}

\documentclass[twocolumn,trackchanges]{aastex631}
\usepackage{tabularx}
\shorttitle{Impact of patchy reionization on dwarfs}
\shortauthors{Kim et al.}

\begin{document}

\title{The Impact of Patchy Reionization on Ultra-faint Dwarf Galaxies}

\correspondingauthor{Myoungwon Jeon}
\email{myjeon@khu.ac.kr}

\author{Jaeeun Kim}
\affiliation{School of Space Research, Kyung Hee University, 1732 Deogyeong-daero, Yongin-si, Gyeonggi-do 17104, Korea}

\author{Myoungwon Jeon}
\affiliation{School of Space Research, Kyung Hee University, 1732 Deogyeong-daero, Yongin-si, Gyeonggi-do 17104, Korea}
\affiliation{Department of Astronomy \& Space Science, Kyung Hee University, 1732 Deogyeong-daero, Yongin-si, Gyeonggi-do 17104, Korea}


\author{Yumi Choi}
\affiliation{National Optical-Infrared Astronomy Research Laboratory (NOIRLab), 950 North Cherry Avenue, Tucson, AZ 85719, USA}
\affiliation{Department of Astronomy, University of California, Berkeley, 501 Campbell Hall, Berkeley, CA 94720, USA}

\author{Hannah Richstein}
\affiliation{Department of Astronomy, University of Virginia, 530 McCormick Road, Charlottesville, VA 22904, USA}

\author{Elena Sacchi}
\affiliation{Leibniz-Institut f{\"u}r Astrophysik Potsdam, An der Sternwarte 16, 14482 Potsdam, Germany}
\affiliation{INAF–Osservatorio di Astrofisica e Scienza dello Spazio di Bologna, Via Gobetti 93/3, I-40129 Bologna, Italy}

\author{Nitya Kallivayalil}
\affiliation{Department of Astronomy, University of Virginia, 530 McCormick Road, Charlottesville, VA 22904, USA}



\begin{abstract}

We investigate how patchy reionization affects the star formation history (SFH) and stellar metallicity of ultra-faint dwarf galaxies (UFDs). Patchy reionization refers to varying ultraviolet (UV) background strengths depending on a galaxy's environment. Recent observations highlight the significance of this effect on UFDs, as UFDs can have different SFHs depending on their relative position with respect to their host halo during the period of reionization. However, most cosmological hydrodynamic simulations do not consider environmental factors such as patchy reionization, and the effect of reionization is typically applied homogeneously. Using a novel approach to implement patchy reionization, we show how SFHs of simulated UFDs can change. Our cosmological hydrodynamic zoom-in simulations focus on UFD analogs with $M_{\rm vir} \sim 10^{9}\msun$, $M_{\rm *} \lesssim 10^{5}\msun$ at $z=0$. We find that patchy reionization can weaken the effect of reionization by two orders of magnitude up to $z=3$, enabling late star formation in half of the simulated UFDs, with quenching times $\sim$460 Myr later than those with homogeneous reionization. We also show that halo merger and mass assembly can affect the SFHs of simulated UFDs, in addition to patchy reionization. The average stellar iron-to-hydrogen ratio, [Fe/H], of the simulated UFDs with patchy reionization increases by 0.22-0.42 dex. Finally, our findings suggest that patchy reionization could be responsible for the extended SFHs of Magellanic UFDs compared to non-Magellanic UFDs.

\end{abstract}



\section{Introduction} \label{sec:intro}

In the context of hierarchical $\Lambda$CDM models of structure formation, where small galaxies are expected to form first and progressively grow into massive ones, it is essential to understand the formation and evolution of low-mass galaxies to obtain a comprehensive understanding of galaxy formation (e.g., \citealp{Padmanabhan1993}). Ultra-faint dwarf galaxies (UFDs), which are known as the smallest unit of galaxies with the lowest luminosity ($L < 10^{5}L_{\rm \sun}$) and stellar mass ($10^{2}\msun < M_{\rm *} < 10^{5}\msun$) in the universe, have been considered as the fundamental building blocks of massive galaxies (reviewed in \citealp{Simon2019}, also see \citealp{Tolstoy2009, McConnachie2012}). Additionally, due to their shallow potential wells, UFD systems are highly vulnerable to internal and external feedback effects, making them an excellent laboratory for studying how feedback mechanisms change star formation activities in small systems (e.g., \citealp{Stinson2007, Sawala2011, Simpson2013, Agertz2015, Jeon2017,  Wheeler2019, Agertz2020, Rey2020, Gutcke2022, Sanati2023}).

The star formation histories (SFHs) of UFDs are generally governed by physical processes such as photoionization heating from stars, supernova feedback (SNe), which releases energy when stars die, and global heating caused by cosmic reionization. Interestingly, a common trait observed in the SFHs of UFD galaxies is that they likely formed the majority of their stars (about 80\%) prior to the onset of reionization, which is followed by a suppression of their star formation. (e.g., \citealp{Brown2014, Weisz2014}). This implies that reionization played a crucial role globally in quenching the star formation of UFD galaxies, along with SN feedback, which dissipates dense gas that is eligible for star formation. Without SN feedback, SFHs of the UFDs can extend to as late as $z\sim2$, otherwise, they are truncated at an early stage around $z\sim6$ (e.g., \citealp{Simpson2013, Jeon2017}). However, despite the importance of cosmic reionization in shaping UFD SFHs, current implementations of reionization in simulations typically adopt a simple spatially uniform UV background model (\citealp{Faucher2009}; \citealp{Madau2012}, hereafter HM2012). To be specific, the UV background intensity used in simulations of UFD analogs is applied uniformly without taking into account the local UV fields originating from their host galaxies.

According to recent research by \citet{Sachhi2021}, the SFHs of UFDs associated with the Magellanic Clouds (MCs) could offer evidence of patchy reionization. To clarify, the MCs consist of the Large Magellanic Cloud (LMC) and the Small Magellanic Cloud (SMC), both being the largest satellite galaxies of the Milky Way (MW). Patchy reionization refers to the non-uniform reionization process, which varies based on the individual environments of UFDs, their proximity to their host galaxy, and the intensity of UV photon emissions from the host halo during the era of reionization. The Magellanic UFDs could present an ideal opportunity to investigate the impact of patchy reionization, primarily because recent data from Gaia DR2 proper motions suggest that they entered the MW’s virial radius relatively recently, less than 3.5 billion years ago (e.g., \citealp{Patel2020}). This implies that Magellanic UFDs might have been situated at a considerable distance from the host halo during the reionization period, leading to a weak influence from the MW halo.

\citet{Sachhi2021} conducted a comparative analysis of the SFHs of Magellanic UFDs, primarily linked to the LMC and having recently entered the MW's halo, with those of long-standing UFD satellites of the MW, which they refer to as non-Magellanic UFDs. Their findings revealed that Magellanic UFDs tend to exhibit more extended SFHs, lasting $\sim$600 Myr, in contrast to non-Magellanic UFDs. Given that both Magellanic and non-Magellanic UFD satellites share similar stellar masses ($M_{\ast}\sim10^{3}\msun$), and a significant fraction of their stars ($\sim80\%$) formed prior to reionization, the discrepancy in their SFHs could be attributed to the effects of patchy reionization. Specifically, Magellanic UFDs may have been located farther from the MW's progenitor halo during the reionization era, experiencing a weaker reionization impact compared to non-Magellanic UFDs, thus explaining their extended SFHs. As such, the effect of reionization on UFDs could vary depending on the environment in which they were located at the time of reionization, leaving an imprint on their SFH. In this study, we investigate how the local UV fields, which are mainly determined by connection with the host halo, can shape the SFHs of satellite UFDs through cosmological hydrodynamic zoom-in simulations.


Several previous studies have investigated how the timing and strength of reionization affect individual dwarf galaxies by adopting uniform UV background radiation (e.g., \citealp{Simpson2013}; \citealp{Bose18}; \citealp{Garrison19}; \citealp{Pereira23}). \citet{Simpson2013}, for example, utilized the table provided by HM2012 by varying the period of reionization, $\Delta z$, during which the intensity of reionization increases from zero to full strength. They discovered that when $\Delta z$ was set to $z=7-6$, the resulting stellar mass of dwarf galaxies with a mass of $M_{\rm vir}=10^9\msun$ at $z=0$ could be larger by one order of magnitude than when $\Delta z$ was set to $z=9-8.9$. It should be mentioned, however, that the HM2012 method may overestimate the ionization emissivity at high redshifts (e.g., $z>10$) by extrapolating the UV background radiation that was derived for lower redshifts. This could lead to premature and excessive heating of the intergalactic medium (IGM) at high redshifts (e.g., \citealp{Puchwein2015}; \citealp{Onorbe2017}), which may significantly suppress star formation in low-mass galaxies like UFD systems.

An alternative approach to implementing the effect of reionization is to solve radiative transfer equations self-consistently to trace UV photons, but this method is computationally expensive and is usually conducted on a large scale with lower mass resolution (e.g., \citealp{Pawlik2008, Dixon18, Rosdahl18}). Such large-volume simulations that have focused on studying the large-scale reionization history, resultant galaxy luminosity functions, and the escape of ionizing radiation have not placed much emphasis on the detailed evolution of small satellite galaxies such as UFDs. Among these, the SPHINX simulation (\citealp{Rosdahl18}) achieved high spatial resolution ($\sim10$ pc) and addressed the radiative transfer aspect in the context of UFD formation.

This paper takes a unique approach to model the influence of reionization on UFDs by applying local UV fields from host galaxies rather than the traditional approach of using a uniform and homogeneous reionization effect. To avoid computationally expensive calculations such as radiative transfer, pre-calculated local UV fields from dark-matter-only simulations are utilized in cosmological hydrodynamic simulations of UFD analogs. In particular, we choose a halo pair consisting of a target UFD and its host halo, analogous to the MW and its satellites, from dark-matter-only simulations. Then, we calculate the strength of local UV fields from the host as a function of the distance between the target UFD and the host and the spectral energy distribution (SED) of the host galaxy.

Furthermore, we test a scenario where reionization happens later than previously thought by incorporating a transition redshift, $z_{\rm t}$, which marks the point at which the impact of the overall UV radiation in the universe becomes stronger than that of the local UV radiation from a host galaxy. Regarding the timing of reionization completion, while it is widely accepted to occur at $z=6$ (e.g., \citealp{Becker2001,Fan2006}), recent spectroscopic studies of Lyman alpha emitters and distant quasars suggest a late reionization scenario, where reionization is completed up to $z=5.5-6$, which potentially leads to prolonged SFHs (\citealp{Becker2015}; \citealp{Choudhury2015}; \citealp{McGreer2015}; \citealp{Mesinger2015}). To better understand whether this delayed reionization can result in extended SFHs in UFD galaxies, we carry out simulations by altering the value of $z_{\rm t}$ from $z=5.8$ to $z=5.5$.


The paper is structured as follows. In Section~\ref{sec:Metho}, we describe the numerical methodology used in this study, while in Section~\ref{sec:result}, we present the simulation results. Our main conclusions are summarized in Section~\ref{sec:conclusion}. Unless stated otherwise, all distances are given in physical units for consistency.

\begin{deluxetable*}{cccccccc}[ht]
\tablecaption{Physical quantities of the simulated UFD analogs at $z=0$ in the GR runs, assuming homogeneous reionization.\label{table1}}
\tabletypesize{\small}
\tablewidth{0pt}
\tablehead{
\colhead{Halo} & \colhead{$M_{\rm vir}$} & \colhead{$r_{\rm vir}$} & \colhead{$M_{\ast}$} & \colhead{$\rm <[Fe/H]>$} & \colhead{$M_{\rm gas}$} & \colhead{$\rm SF_{\rm trun}$} & \colhead{$z_{\rm sf,end}$} \\
\colhead{} & \colhead{$[10^9\msun]$} & \colhead{$[\rm kpc]$} & \colhead{$[10^5\msun]$} & \colhead{-} & \colhead{$[10^6\msun]$} & \colhead{-} & \colhead{-}
}
\startdata
{\sc Halo1-GR}  & 0.74 &  18.5  & 0.14 & -3.28  & 1.5 & yes & 7.036 \\
\hline
{\sc Halo2-GR}  & 1.02 &  20.8  & 0.38 & -2.43 & 2.0 & yes & 6.998 \\
\hline
{\sc Halo3-GR}  & 1.05 & 20.7 & 0.24 & -2.81 & 2.6 & yes & 7.106 \\
\hline
{\sc Halo4-GR}  & 1.09 &  21.0  & 0.14 & -2.82 & 3.2 & yes & 7.037 \\
\hline
{\sc Halo5-GR}  & 1.40 &  22.9  & 0.63 & -2.33 & 2.0 & yes & 6.917 \\
\hline
{\sc Halo6-GR}  & 1.84 & 25.0  & 0.50 & -2.37 & 7.0 & yes & 6.895 \\
\enddata
\tablecomments{Column (1): the name of halos. Column (2): viral mass (in units of $10^9\msun$).
Column (3): virial radius (in kpc). Column (4): stellar mass (in $10^5\msun$). Column (5): average stellar iron-to-hydrogen ratios. Column (6): gas mass (in $10^6\msun$). Column (7): whether star formation is truncated after reionization. Column (8): the time when star formation is completed.}
\end{deluxetable*}

\section{Numerical methodology} \label{sec:Metho}
\subsection{Simulation Set Up} \label{sec:Metho_setup}

We have utilized a modified version of the N-body and Smoothed Particle Hydrodynamics (SPH) code GADGET (\citealp{Springel2001}; \citealp{Springel2005}) to perform a set of hydrodynamic zoom-in simulations. The adopted cosmological parameters include a matter density parameter of $\Omega_{\rm m}=1-\Omega_{\Lambda}=0.265$, a baryon density of $\Omega_{\rm b}=0.0448$, a present-day Hubble expansion rate of $\rm H_0=71\unit{km\enspace s^{-1}\enspace Mpc^{-1}}$, a spectral index of $n_{\rm s}=0.963$, and a normalization of $\sigma_8=0.8$ (\citealp{Komatsu2011,planck2016}). The initial conditions are generated using the cosmological initial conditions code MUSIC (\citealp{Hahn2011}). A preliminary dark-matter-only simulation using $128^3$ particles in a $L=6.25 h^{-1} \cMpc$ box is carried out to select target halos. The target halos chosen represent UFD analogs with a mass of $M_{\rm vir} \sim 10^9\msun$ at $z=0$, and their physical properties are listed in Table~\ref{table1}. It is important to emphasize that we choose UFD analogs that are isolated and situated at a considerable distance from the host halo throughout their evolution. Our primary purpose is to study how patchy reionization affects the SFHs of these UFD analogs, particularly during high-$z$. By focusing on isolated UFD analogs, we can exclude the potential impact of environmental factors, such as ram pressure stripping, as they come closer to the MW.

Next, we have conducted four consecutive refinements to the region enclosing the area two times the virial radius of the target UFD halo at $z=0$. In the most refined region, the final resolution is $2048^3$, with dark matter (DM) and gas-particle masses of $m_{\rm DM}\approx2500\msun$ and $m_{\rm SPH}\approx500\msun$, respectively. The softening length of DM and stellar particles is fixed at 40 pc at all redshifts in our simulations. However, for gas particles, we utilize an adaptive softening length that is proportional to the minimum value of $\epsilon_{\rm gas, min}=2.8$ pc. At each time step, we solve the non-equilibrium rate equations for the primordial chemistry of nine atomic and molecular species (H, H+, H-, H2, H+2, He, He+, He++, e-, D, D+). Besides primordial cooling, we have incorporated metal cooling processes with carbon, oxygen, silicon, magnesium, neon, nitrogen, and iron. The cooling rates for these elements in the simulation are determined using the photo-ionization package CLOUDY (\citealp{Ferland1998}).

\subsection{UV background} \label{sec:Metho_uv}
\subsubsection{Global reionization} \label{sec:Metho_uv_gl}

This section describes one of two methods used to implement the UV background in our simulations. There are two distinct approaches we have utilized for this purpose. The first method is the homogeneous and flash-like cosmic UV/X-ray background, which was proposed by \citet{Madau2012}. This approach involves the use of redshift-dependent photoionization and photoheating rates for H I, He I, and He II to mimic the process of reionization. This effect is uniformly applied to all galaxies within the simulation box, regardless of their position relative to their host galaxy. From this point on, we will refer to the homogeneous reionization effect as global reionization (GR). In our simulations, we start by introducing the UV background at $z=7$ and progressively increase it to the full strength at $z=6$ to prevent abrupt and significant heating of gas particles. Then, we maintain it at the constant 100\% strength of HM2012 until $z=0$.

\subsubsection{Patchy reionization} \label{sec:Metho_uv_pr}
The second approach is a patchy UV background, which takes into account environmental factors specific to each target UFD halo, such as the proximity of ionizing sources and the distance between them, which may alter the reionization effect experienced by the target halo. To achieve the most accurate results in calculating the effect of patchy reionization (PR) for our UFD analogs, it would be ideal to track the trajectory of ionizing photons emitted from the surrounding galaxies and apply the resulting heating and ionization effects on the target dwarf galaxy. However, carrying out a hydrodynamic simulation that self-consistently solves radiative transfer equations necessitates a significant amount of computational power. 

To overcome this computational challenge, we have devised a novel approach where we pre-calculate the impact of local UV fields generated by the surrounding galaxies using DM-only simulations. This pre-calculated information is then utilized in our cosmological hydrodynamic simulations. In summary, obtaining pre-calculated UV fields involves three steps. Firstly, a DM-only simulation is performed to determine the environmental factors, such as the distance between a target UFD analog and the surrounding DM halos, as well as their halo masses. Secondly, we employ the abundance-matching approach proposed by \citet{Behroozi2013} to estimate the stellar mass of the surrounding halos. Finally, to attain the photoionization and photoheating rates from the neighboring galaxies, we use the Starburst99 package (\citealp{Leitherer1999}) to derive the SED. Below, we provide a detailed explanation for each step.

\begin{figure*}
  \centering    
  \includegraphics[width=165mm]{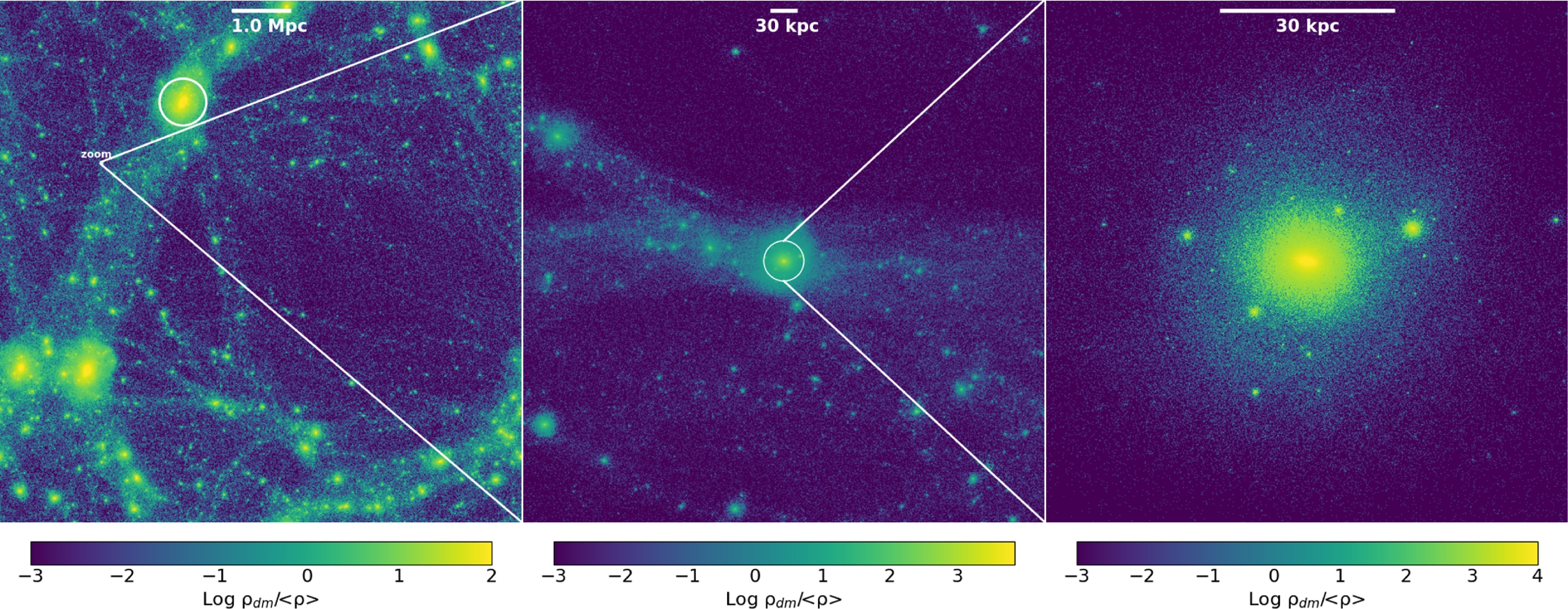}%
   \caption{Dark matter morphology at $z=0$, with each panel depicting a smaller scale from left to right. {\sffamily Left:} final snapshot from a dark-matter-only preliminary run that includes an MW-like halo, marked as a large white circle, and a UFD analog with masses of $M_{\rm vir}\sim2\times10^{12}\msun$ and $M_{\rm vir}\sim10^{9}\msun$ at $z=0$, respectively. {\sffamily Middle:} the zoom-in of the region surrounding the UFD analogs. {\sffamily Right:} detailed look of {\sc Halo6-GR} (see Table~\ref{table1}). The virial radius of the halo is $\sim$ 30 kpc at $z=0$.}
   \label{fig1}
\end{figure*}

The first step of the process includes carrying out a dark-matter-only simulation with $128^3$ particles using the same simulation setup and cosmology as described in Section~\ref{sec:Metho_setup}. Following this, a target UFD analog is selected, and the distance between the target UFD analog and the surrounding halos, including its host halo, as well as the mass evolution of the surrounding halos, are tracked. Figure~\ref{fig1} displays the DM morphology at $z=0$, which highlights a pair consisting of the target UFD ($M_{\rm vir}\sim10^9\msun$) and its host halo ($M_{\rm vir}\sim2\times10^{12}\msun$). 

In the second step, we estimate the stellar masses of the surrounding halos using the abundance matching technique (\citealp{Behroozi2013}). This is a widely used approach for assigning galaxy stellar mass or luminosity to halos generated in N-body simulations without having to perform hydrodynamical simulations.

In the final step, we adopt the Starburst99 package to derive the SED of the simulated galaxies. To do this, we utilize standard parameters, employing the Geneva evolutionary tracks without rotation as a model for stellar evolution, and black body spectra as a model for the atmosphere. Starburst99 generates a synthetic spectrum of galaxies based on their stellar masses, which were obtained in the previous step using the abundance matching technique. Since the massive OB stars formed in the neighboring galaxies are the primary source of ionizing photons, and their main sequence lifetime is approximately 10 Myr, we divide the stellar mass evolution of the surrounding galaxies into time intervals that are similar to the lifetime of OB stars. We then estimate the change in stellar mass ($\Delta M_{\rm *}$) due to newly formed stars in the surrounding galaxies. We calculate the local UV fields exerted by the neighboring galaxies onto the target UFD analog by plugging in the distance between them and the synthetic spectrum of the surrounding galaxies into equations (1) and (2).

The rates of photoionization and photoheating are given by

\begin{equation}
\rm K_{\rm ion,i}=\int_{\nu_{\rm min}}^{\infty}\frac{F_{\rm \nu}\sigma_{\rm \nu}} {\emph{h}\nu}\, d\nu,
\end{equation}

\begin{equation}
\Gamma_{\rm i}=n_{\rm n}\int_{\nu_{\rm min}}^{\infty}{\rm F_{\rm \nu}}\sigma_{\rm \nu}(1-\frac{\nu_{\rm min}} {\nu})\, d\nu. 
\end{equation}

Here, $\sigma_{\nu}$ refers to the ionization cross-section (\citealp{Osterbrock2006}), $n_{\rm n}$ is the number density of the respective neutral species, while $\nu_{\rm min}$ represents the ionization threshold frequency, such as $h\nu_{\rm min}=13.6$ eV, $h\nu_{\rm min}=24.6$ eV, and $h\nu_{\rm min}=54.4$ eV for H~I, He~I, and He~II, respectively. The ionizing flux, $\rm F_{\rm \nu}$, incident upon the target UFD analog, originating from the surrounding galaxies, is computed by taking into account the distance between the target halo and the neighboring galaxies and their SEDs. Self-shielding of the dense gas is implemented by attenuating the UV background based on $\exp{(-N_{\rm HI} \bar{\sigma}_{\rm ion}})$, where $N_{\rm HI}=x n_{\rm HI}$ (\citealp{Simpson2013}). Here, $x$ represents the SPH kernel size, $n_{\rm HI}$ denotes the neutral hydrogen number density, and $\bar{\sigma}_{\rm ion}$ refers to the frequency-averaged photoionization cross-section for HI. Similarly, a simple approach to self-shielding is applied for He~I and He~II as well.

It should be noted that the halo masses of the surrounding halos range from approximately $M_{\rm vir}\approx 10^{10}\msun$ to $M_{\rm vir}\approx 10^{12}\msun$ at $z=0$. While we consider roughly 40 surrounding galaxies, including a host galaxy within the simulated box, we find that the impact of the host halo on the target UFD analog is likely to outweigh the effects of other smaller halos. Therefore, we will only focus on the influence of the host halo going forward. Additionally, we conduct simulations to examine whether the trajectories of UFD analogs in relation to the MW-like host halo remain consistent between simulations with hydrodynamics and the DM-only simulation. We verify that the difference in distance between the two sets of simulations is less than 1\%, indicating that the DM-only simulation aligns with the hydro simulations in terms of proximity to the host halo.

\subsubsection{Transition time, $z_{\rm t}$} \label{sec:Metho_uv_zt}
In the PR implementation discussed in the preceding section, we only take star-forming galaxies into consideration as the ionizing sources of reionization. However, HM2012 suggests that hard UV-emitting quasars should also be considered as a source of reionization alongside star-forming galaxies. The precise contribution of these two ionizing sources as a function of redshift remains uncertain, although it is anticipated that quasars will be the dominant source at lower redshifts ($z<5$) (e.g., \citealp{Wyite2003}; \citealp{Madau2012}).To take into account the influence of quasars on the reionization process, which is beyond the scope of this study, we combine our patchy UV background approach with the homogeneous reionization provided by HM2012. To be specific, we apply our patchy reionization implementation at relatively higher redshifts, where star-forming galaxies are the primary source of reionization and the impact of quasars is insignificant. We then designate a transition time, $z_{\rm t}$, where we switch from the patchy reionization approach to the homogeneous reionization model proposed by HM2012. 

The transition period, ranging from $z=5.5$ to $z=5.8$, is inspired by late reionization scenarios, which suggest that the end of reionization could extend from $z=6$ to as late as $z=5.5$, as proposed in various studies (\citealp{Becker2015}; \citealp{Choudhury2015}; \citealp{McGreer2015}; \citealp{Mesinger2015}). Given the significant impact of GR, such that star formation in UFD analogs is immediately suppressed upon its introduction, the transition time can be regarded as marking the end of reionization. Indeed, this combined approach enables an exploration of how patchy reionization might impact the SFHs of the simulated UFD analogs and provides insights into when reionization reaches its completion.

Moreover, we incorporate an escape fraction, $f_{\rm esc}$, as a free parameter, which represents the fraction of ionizing photons from star-forming galaxies that can escape and propagate into the IGM. The estimate of $f_{\rm esc}$ varies with halo mass and redshift, and theoretical studies have estimated the average value of $<f_{\rm esc}>$ to be 0.1-0.2 at high redshifts ($z\gtrsim5$) (e.g., \citealp{Kimm2014}; \citealp{Bouwens15b}; \citealp{Mitra15}; \citealp{Khaire16}; \citealp{Ma2020}), with some studies suggesting a somewhat higher value of $<f_{\rm esc}> \sim0.4$ (e.g., \citealp{Yajima11}). Increasing the escape fraction has a stronger impact on reionization, resulting in the heating of gas within low-mass halos and the consequent suppression of star formation in such systems. Observational studies have extensively investigated the constraints on the escape fraction, particularly for galaxies at lower redshifts (e.g., \citealp{Choi2020}; \citealp{Mestric2021}; \citealp{Naidu2022}). Notably, \citet{Choi2020} proposed a value of $\sim$0.25 for $f_{\rm esc}$ based on the SED analysis of resolved stars in NGC 4214. Despite being a dwarf galaxy at a low redshift, NGC 4214 exhibits properties that make it a suitable analog for studying the ionizing sources responsible for reionization.

The $f_{\rm esc}$ can also be influenced by the binary population within galaxies (e.g., \citealp{Ma2016}; \citealp{Rosdahl18}). In particular, \citet{Rosdahl18}, who investigated the impact of binary population on $f_{\rm esc}$ and the reionization history, proposed that considering binary stars leads to about three times higher $f_{\rm esc}$ in observed 1500$\Ang$ than a single stellar population. However, due to the challenge of constraining the frequency of binary stars, we have not accounted for the effect of binaries in our SED models, which would yield harder resulting SEDs. \citet{Choi2020} also did not incorporate stellar rotation or binary evolution in their SED models. In our simulations involving patchy UV fields, we adopt a fixed value of $f_{\rm esc}=0.3$ for the range $z_{\rm t} < z < 7$, considering that star-forming galaxies are the primary contributors to reionization during this epoch. This choice is reasonable for the host galaxies involved in the reionization process.

\subsection{Star formation} \label{sec:Metho_sf}
We include the formation of Population III (Pop III) stars, metal-free first-generation stars, and Population II (Pop II) stars, which form from gas clouds enriched by metals from the SN explosions of Pop III stars. The transition from Pop III to Pop II star formation occurs when the metallicity of the gas cloud exceeds a critical value of $Z_{\rm crit}=10^{-5.5}\zsun$ (\citealp{Omukai2000, Schneider2010, Safranek2016}). For a detailed description of the star formation recipe and associated stellar feedback, we refer readers to \citet{Jeon2017}. In short, star formation is triggered when the hydrogen number density of a gas particle surpasses the threshold value of $n_{\rm H,th}=100 \cmci$. The gas particle is transformed into collisionless star particles with a mass of $m_{\rm star}=500\msun$. Instead of treating each star particle as an individual star, we consider them as a single stellar cluster. We assume that the initial mass function (IMF) for Pop III stars is top-heavy, $\phi_{\rm PopIII}(m) = dN/d\log m\approx m^{-\alpha}$, with a slope $\alpha=1.0$ and covers the mass range of $[m_0,m_1]=[10,150]\msun$. For Pop II stars, we adopt the Chabrier IMF within the mass range of $[m_0,m_1]=[0.1, 100]\msun$. 

We implement a stochastic conversion of gas particles to star particles based on Schmidt's law (\citealp{Schmidt1959}), where stars are formed at a rate of $\dot{\rho}{\ast}=\rho/\tau_{\ast}$. Here, $\rho$ is the gas density, and the star formation timescale, $\tau_{\ast}$, is given by $\tau_{\ast} = \tau_{\rm ff} / \epsilon_{\rm ff}$, where $\tau_{\rm ff}$ corresponds to the free-fall time, and $\epsilon_{\rm ff}$ denotes the star formation efficiency per free-fall time. During each numerical time interval of $\Delta t$, the conversion from gas to star particle only occurs when a randomly generated number between 0 and 1 is less than the minimum of $\Delta t/\tau_{\rm *}$ and 1. We set the star formation efficiency, $\epsilon_{\rm ff}\sim0.01$, for both Pop III and Pop II stars, and the star formation timescale is given by

\begin{equation}
\tau_{\ast}=\frac{\tau_{\rm ff} (n_{\rm H})}{\epsilon_{\rm ff}}\sim400 {\rm Myr} \left(\frac{n_{\rm H}}{100 \cmci}\right)^{-1/2},
\end{equation}
where the free fall time is $\tau_{\rm ff}=[3\pi/(32G\rho_{ \rm})]^{1/2}$. The level of star formation activity in galaxies is regulated by various factors, including their mass and the potency of associated SN feedback in suppressing subsequent star formation. This interplay between galaxy mass and the effectiveness of SN feedback can lead to diverse SFHs, largely characterized by either continuous or bursty star formation.

It is important to highlight that in this study, we have not considered the impact of photoionization heating from Pop~III and Pop~II stars, primarily due to the considerable computational cost. The influence of radiative feedback from these local sources remains a topic of ongoing debate. For instance, \citet{Hopkins2020} suggests that the stellar mass of small galaxies (with a halo mass of $M_{\rm vir}=2-4\times10^9\msun$) is predominantly shaped by external UVB radiation, while local sources have a negligible effect. Conversely, \citet{Agertz2020} have presented contrasting results, suggesting that photoionization heating from stars could reduce the stellar mass by a factor of 5-10 for dwarf galaxies with a halo mass of $M_{\rm vir} = 10^9 \msun$.

\subsection{Chemical feedback} \label{sec:Metho_chem}
We account for chemical enrichment through the contribution of winds from asymptotic giant branch (AGB) stars and the explosions of CCSNe and Type Ia SNe, following the implementation described in \citet{Wiersma2009b}. At each simulation time step, we estimate the masses of 11 individual elements produced by dying stars and release them into the neighboring medium. These elements undergo diffusive mixing in both the interstellar medium (ISM) and the IGM. The initial masses of Pop III stars determine their nucleosynthetic yields and remnant masses. For instance, Pop III stars with initial masses between $10\msun$ and $40\msun$ end their lives in core-collapse supernovae (CCSNe), while those with masses between $140\msun$ and $260\msun$ end their lives in pair-instability supernovae (PISNe). We adopt the nucleosynthetic yields and remnant masses for CCSNe of Pop III stars from \citet{Heger2010} and for PISNe from \citet{Heger2002}.

Pop II stars undergo mass loss through AGB or SN in their final stage. We incorporate metallicity-dependent tables ranging from $Z=0.0004$ to $Z=1.0$ (\citealp{Portinari1998}) to determine the yield and evolution of these stars. Intermediate mass stars ($0.8\msun\lesssim m_{\rm *}\lesssim8\msun$) can lose up to 60\% of their mass during the terminal AGB stage, and these yields are taken from \citet{Marigo2001}. CCSNe from massive stars ($m_{\rm *}\gtrsim8\msun$) release significant amounts of metals, and Type Ia SNe are expected to occur for stars with masses in the range $3\msun\lesssim m_{\rm *}\lesssim8\msun$. Due to uncertainties in the detailed evolution of Type Ia SNe, we use an empirical delay time function expressed in terms of e-folding times (e.g., \citealp{Barris2006}; \citealp{Forster2006}). Metals from dying Pop~II stars are also transported to the IGM and ISM via diffusion-based metallicity, as described by \citet{Greif2009}, where ejected metals disperse into neighboring gas particles ($N_{\rm ngb}=48$). The initial metallicity of the surrounding gas is given below.

\begin{equation}
Z_{\rm i} = \frac{m_{\rm metal,i}} {m_{\rm SPH}+ m_{\rm metal,i}},
\end{equation}
where $m_{\rm metal,i}$ represents the mass of metal assigned to one of the neighboring gas particles, and $m_{\rm SPH}$ is the mass of a gas particle.

\begin{figure*}
  \centering
  \includegraphics[width=140mm]{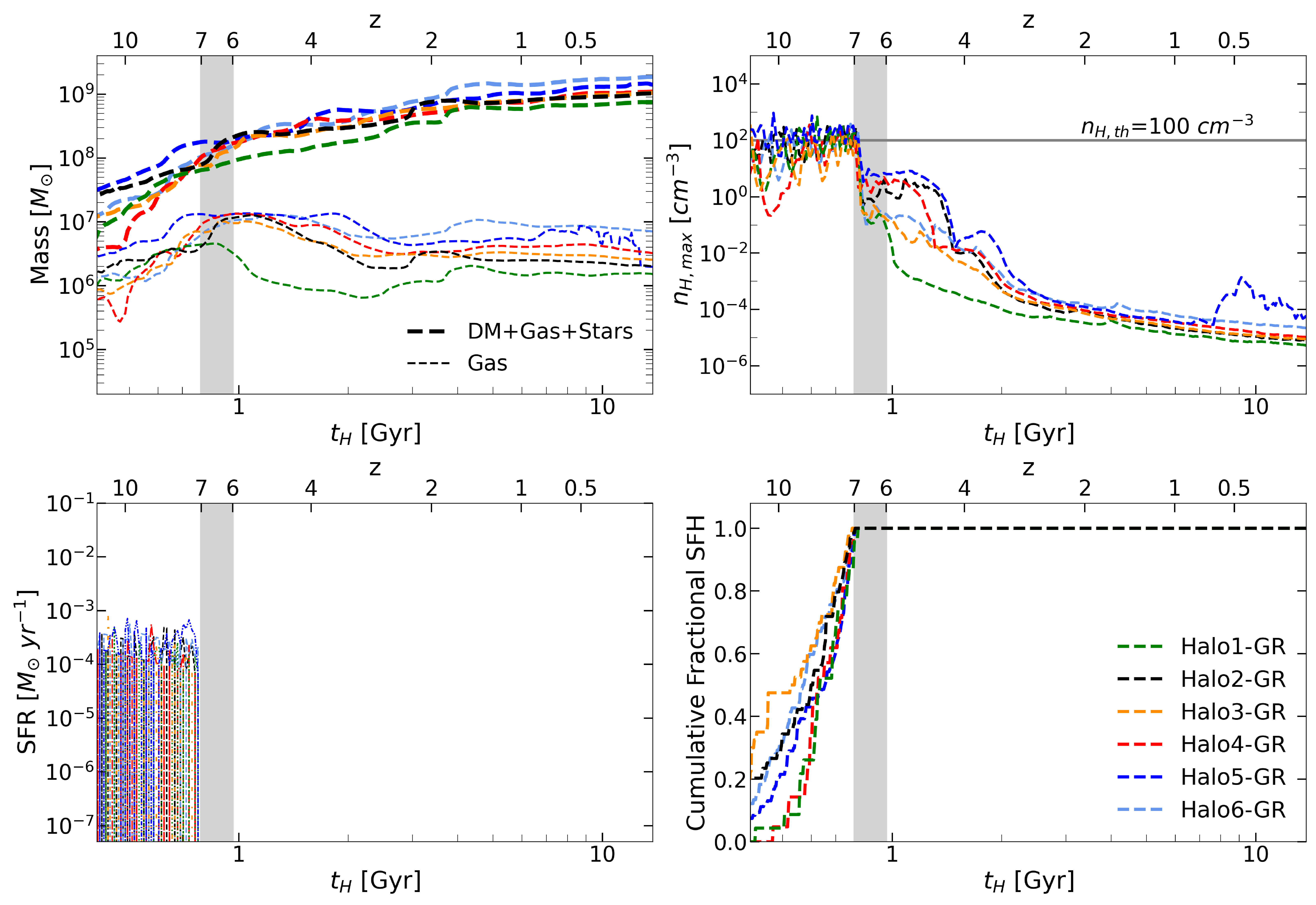}
   \caption{The evolution of the simulated UFD analogs in the GR runs. {\sffamily upper-left:} the total mass (long-dashed line) and gas mass (short-dashed line) of the UFD analogs. {\sffamily upper-right:} the maximum hydrogen number density. {\sffamily lower-left:} the star formation rate. {\sffamily lower-right:} the cumulative star formation history. The quantities presented in all panels are computed using particles residing within the virial radius of the halos. The grey-shaded regions in each panel denote the period when a homogeneous UV background field is introduced at $z=7$, with its strength gradually increasing up to the full value by $z=6$. The evolution of each UFD analog realization is depicted using different colors and line types. Due to the effects of reionization and SN feedback, all simulated galaxies experience truncated star formation. The upper-right panel shows that there is no dense gas above the density threshold, $n_{\rm H, th}=100 \cmci$, suggesting that no further star formation has occurred since the epoch of reionization.}
   \label{fig2}
\end{figure*}

\subsection{Thermal feedback} \label{sec:Metho_thermal}
When a star reaches the end of its life and undergoes an SN explosion, we release the energy from the explosion to neighboring gas particles as thermal energy. However, it is well known that an over-cooling problem can arise if the SN energy is deposited onto too much surrounding gas, resulting in the thermal energy being radiated away (e.g., \citealp{Stinson2007}). To avoid the over-cooling problem associated with SN explosions, we use the method suggested by \citet{Vecchia2012}, where a temperature increase of more than $10^{7.5}$ K is guaranteed by limiting the number of neighboring particles that receive SN thermal energy. To achieve this, we deposit the energy from the SN explosion onto a single neighboring particle ($N_{\rm nbg}=1$) to ensure that the effect of the explosion is preserved. 
The total SN energy per unit solar mass, $\epsilon_{\rm SN}$, is calculated using the adopted IMF for Pop III and Pop II stars, assuming that each SN releases $10^{51}$ erg of energy. This is expressed as $\epsilon_{\rm SN}=n_{\rm SN}\times10^{51}$, where $n_{\rm SN}$ is the number of SNe per unit mass. $n_{\rm SN}$ is obtained by integrating the IMF, $\phi(m)$, over the mass range from $m_0$ to $m_1$. Here, $m_0=8\msun$ and $m_1=100\msun$ are the minimum and maximum initial mass of stars that can undergo SN, respectively. For Pop III stars, the resultant value is $\epsilon_{\rm SN, PopIII}=5.56\times10^{49}$ erg ${\msun}^{-1}$, while for Pop II stars, it is $\epsilon_{\rm SN, PopII}=1.73\times10^{49}$ erg ${\msun}^{-1}$.

\section{Results} \label{sec:result}
In this section, we present the results of our simulations, focusing on how patchy reionization affects the SFHs and stellar metallicities of our simulated UFD analogs. Specifically, we compare two scenarios: one with a homogeneous UV background throughout the entire cosmic history until $z=0$, and the other incorporating patchy reionization effects on the galaxy analogs. Section~\ref{sec:result_gl} examines the fundamental properties of our simulated UFD analogs with a homogeneous UV background. Section~\ref{sec:result_pr} investigates the impact of patchy reionization on the SFHs of our simulated galaxies. Section~\ref{sec:result_zt} analyzes how the duration of star formation varies depending on the transition time, $z_{\rm t}$, from patchy to homogeneous reionization. Finally, we compare the physical properties of our simulated UFDs with those of observed UFDs in the MW and discuss the implications of patchy reionization on UFDs in Section~\ref{sec:result_obs}

\subsection{Basic properties (GR)} \label{sec:result_gl}
Figure~\ref{fig2} exhibits the evolution of the simulated UFD analogs using a homogeneous UV background from the initial star formation activities until $z=0$. The basic properties of the simulated galaxies at $z=0$ are provided in Table~\ref{table1}. The four panels, arranged clockwise from the upper left, illustrate the mass assembly for the virial and gas masses, the maximum hydrogen number density, the cumulative SFH, and the star formation rate (SFR). All physical quantities in each panel are calculated using the particles found within the virial radius of the halo at a given time. A grey-shaded region in each panel denotes the reionization period, during which a homogeneous UV background is introduced at $z=7$, gradually ramping up to its full value by $z=6$. Notably, all the simulated analogs have a virial mass of $M_{\rm vir} \approx$ 1-2 $\times 10^9\msun$ at $z=0$.

We find that the total gas mass within the halos tends to decrease after the UV background reaches its full strength, and the degree of gas loss due to reionization depends on how massive a halo is when the onset of reionization is initiated ($z=7$). As shown in the upper-left panel of Figure~\ref{fig2}, four sets of halos ({\sc Halo1-GR, Halo2-GR, Halo3-GR, Halo4-GR}) attain gas masses of $\sim5\times10^{6}$$\msun$ at $z=7$. In contrast, {\sc Halo5-GR}, the most massive set of halos at $z=7$, having a total mass of $\sim$ 1.6 $\times$ 10$^{8}$$\msun$, tends to preserve a gas mass of $M_{\rm gas}\sim$ 1.3 $\times$ 10$^{7}$$\msun$, which is three times larger than those of the other four halos. On the other hand, {\sc Halo2-GR} loses roughly 90\% of its gas mass ($M_{\rm gas}\sim2\times10^{6}\msun$ at $z=0$) between $z=6$ and $z=0$.

In addition to the total gas mass, the process of reionization also disperses dense gas clouds within the halos. The upper-right panel of Figure~\ref{fig2} displays the maximum hydrogen number density of gas particles within each halo, compared to the density threshold for star formation, $n_{\rm H,th}=100 \cmci$, represented by a solid gray horizontal line. Gas particles above this density threshold can be transformed into star particles with a mass of $m_{\rm star}$. The maximum gas densities of all halos decrease significantly by two orders of magnitude from $z=7$ to $z=6$. The UV background effectively disperses the dense gas particles, making it difficult to form new stars, resulting in no star formation in all halos below $z\approx7$. Interestingly, {\sc Halo2-GR, Halo4-GR}, and {\sc Halo5-GR} can retain the gas particles with $n_{\rm H}\sim 1-10 \cmci$ for $\sim$ 250 Myr at $z\lesssim6$. This is due to the relatively higher virial masses of these halos, with $M_{\rm vir}\sim2.0\times10^{8}\msun$, $\sim1.6\times10^{8}\msun$, and $\sim1.8\times10^{8}\msun$ for {\sc Halo2-GR, Halo4-GR} and {\sc Halo5-GR}, respectively, at $z=6$. On the other hand, {\sc Halo1-GR}, which has the lowest mass with $M_{\rm vir}\sim8.6\times10^{7}\msun$ at $z=6$, shows a significant reduction in the maximum hydrogen number density by five orders of magnitude between $z=7$ and $z=6$.

\begin{figure}
  \centering
  \includegraphics[width=85mm]{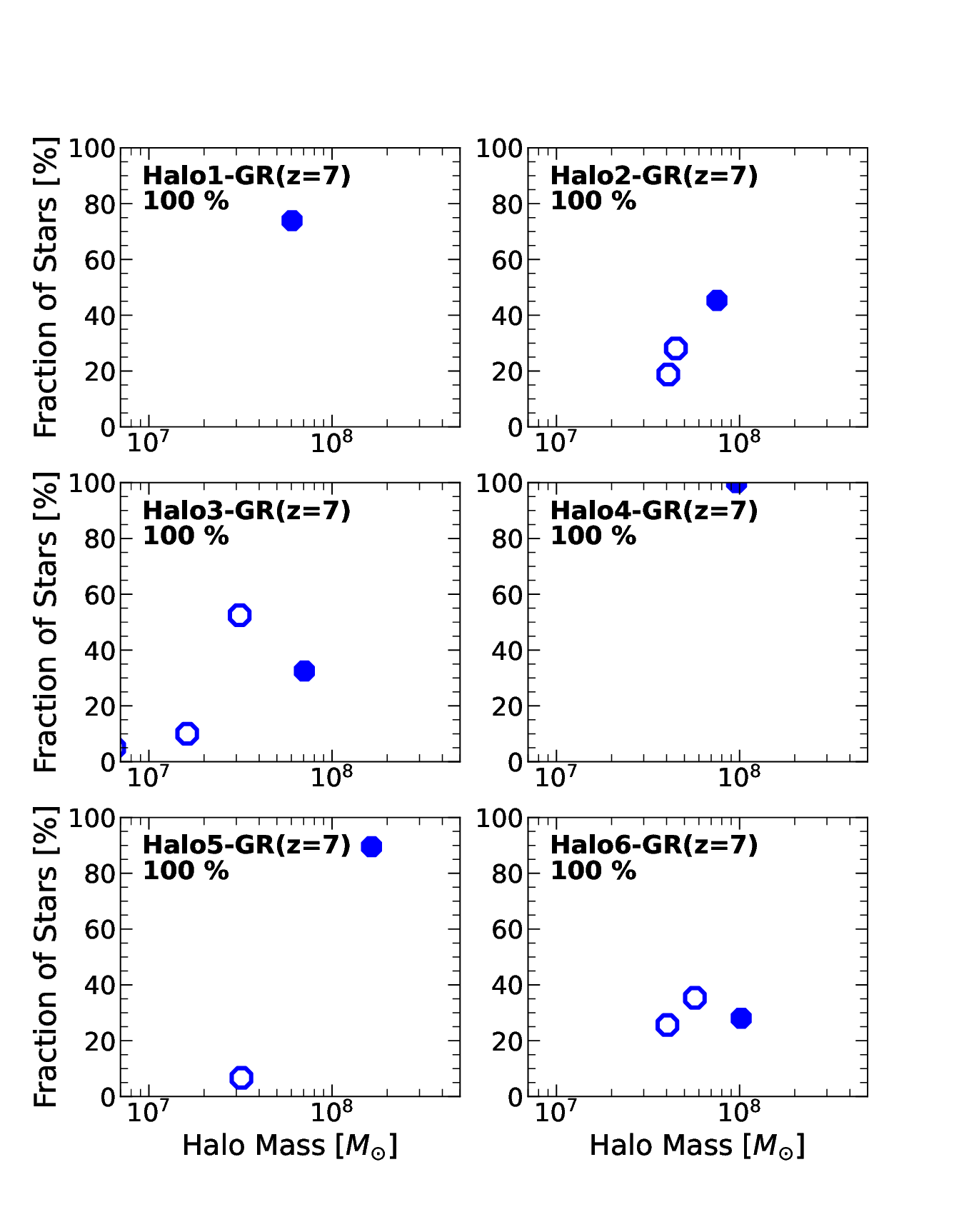}
   \caption{The fraction of stars formed within the progenitor halos of the UFD analog is shown in each panel as a function of halo masses at $z=7$. Only progenitor halos that have formed stars contributing to more than 5\% of the stellar mass of the UFD analog at $z=0$ are displayed. The value in the upper corner of each panel represents the proportion of stars formed at $z=7$ compared to the total number of stars formed by $z=0$, implying that all stars formed at $z=7$. The primary halo, the most massive halo at a given epoch, is depicted as a filled blue circle. This suggests that the number of progenitor halos contributing to the total stellar mass can vary depending on how fast the primary halo grows.}%
   \label{fig3} 
\end{figure}

This quenching trend by reionization is reflected as a truncated SFH in the bottom-right panel of Figure~\ref{fig2}. The SFHs in this panel represent the cumulative fraction of stars formed until a given time among all stars in each halo at $z=0$. For instance, all simulated halos exhibit a ratio of unity at $z\sim7$, implying that all stars in each halo are formed prior to reionization. Our findings are consistent with previous studies (e.g., \citealp{Brown2014, Weisz2014, Jeon2017}), confirming that low-mass progenitor halos of the UFD analogs are vulnerable to internal and external processes, such as SN feedback and reionization, giving rise to short SFHs.

It should be emphasized that the simulated galaxies are the result of the merging of multiple progenitor halos, indicating that they have several small-mass progenitors at high redshifts. Figure~\ref{fig3} illustrates the progenitor halos, which are the halos where the stars that are found within the virial radius at $z=0$ are formed and then merged with the primary halo. Only the progenitors that contribute more than 5\% of the final stellar mass of the simulated galaxies are shown, with a filled circle representing the primary halo, the most massive halo among the progenitors. The percentage of the fraction of stellar mass at $z=7$ in comparison to the final stellar mass of the UFD analogs at $z=0$ is presented below the halo name in each panel of Figure~\ref{fig3}. The percentage is 100\% in all panels, which is due to the truncated SFHs at $z=7$.

\begin{figure}
  \centering
  \includegraphics[width=84mm]{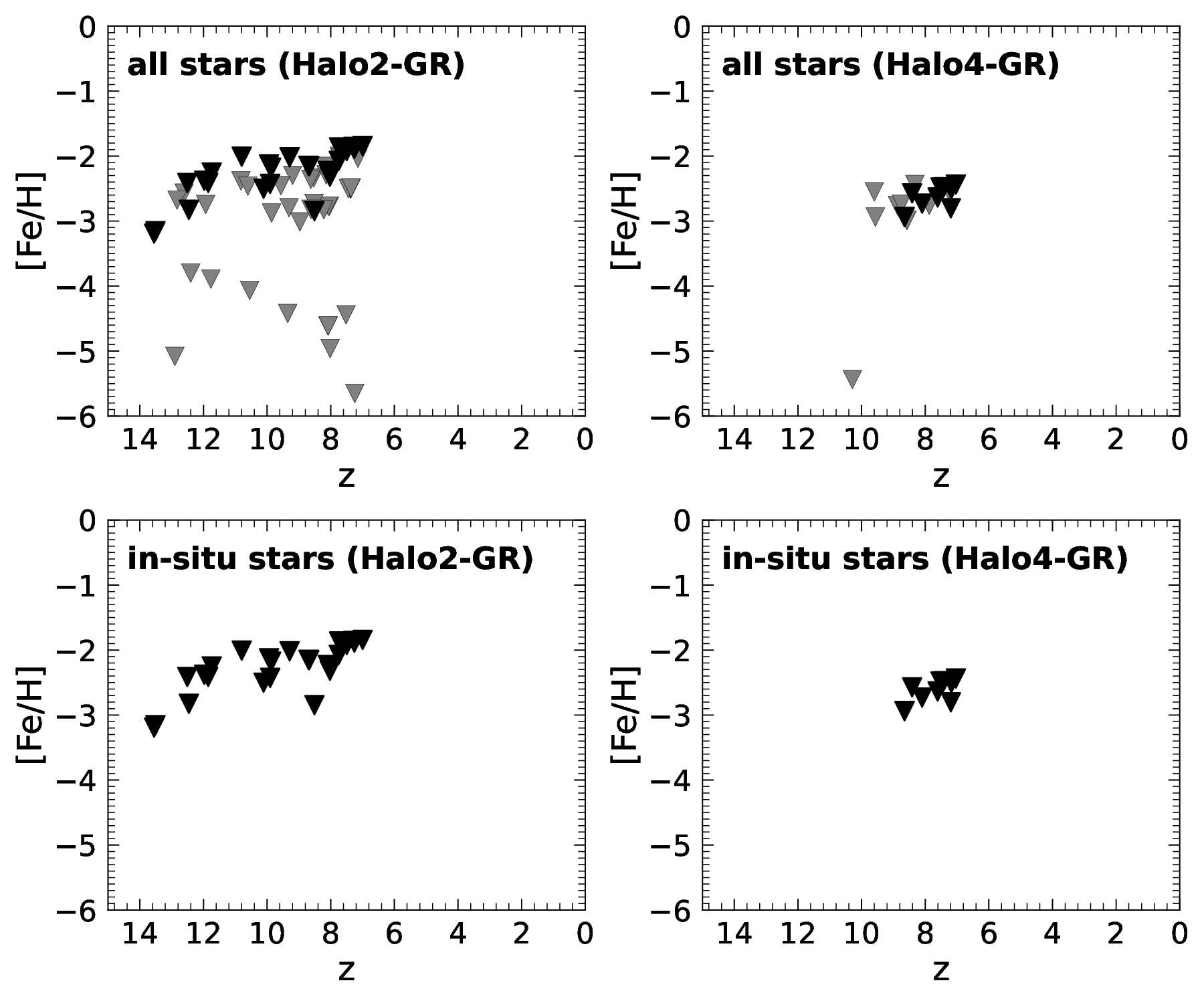}
   \caption{The metallicity evolution of stars as a function of their formation time in {\sc Halo2-GR} (left panel) and {\sc Halo4-GR} (right panel). All the stars found within the virial radius of the halos at $z=0$ are shown in the upper panels, while the bottom panels show only the in-situ stars in the primary progenitor of a halo. Among all the stars in the upper panels, we differentiate the in-situ stars by coloring them in black. As depicted in the run, {\sc Halo2-GR}, stars with relatively low metallicity ($\rm [Fe/H]<-2$) usually form in progenitor halos that later merge with the primary halo. The evolution of the $\rm [Fe/H]$ in the primary halo shows a rising trend with redshift, as seen in the case of {\sc Halo2-GR}. At $z\approx13$, the value of $\rm [Fe/H]$ is $\rm [Fe/H]\approx-3$, which increases to $\rm [Fe/H]\approx-2$ at $z\approx6$.}
   \label{fig4} 
\end{figure}

\begin{figure*}
  \centering
  \includegraphics[width=145mm]{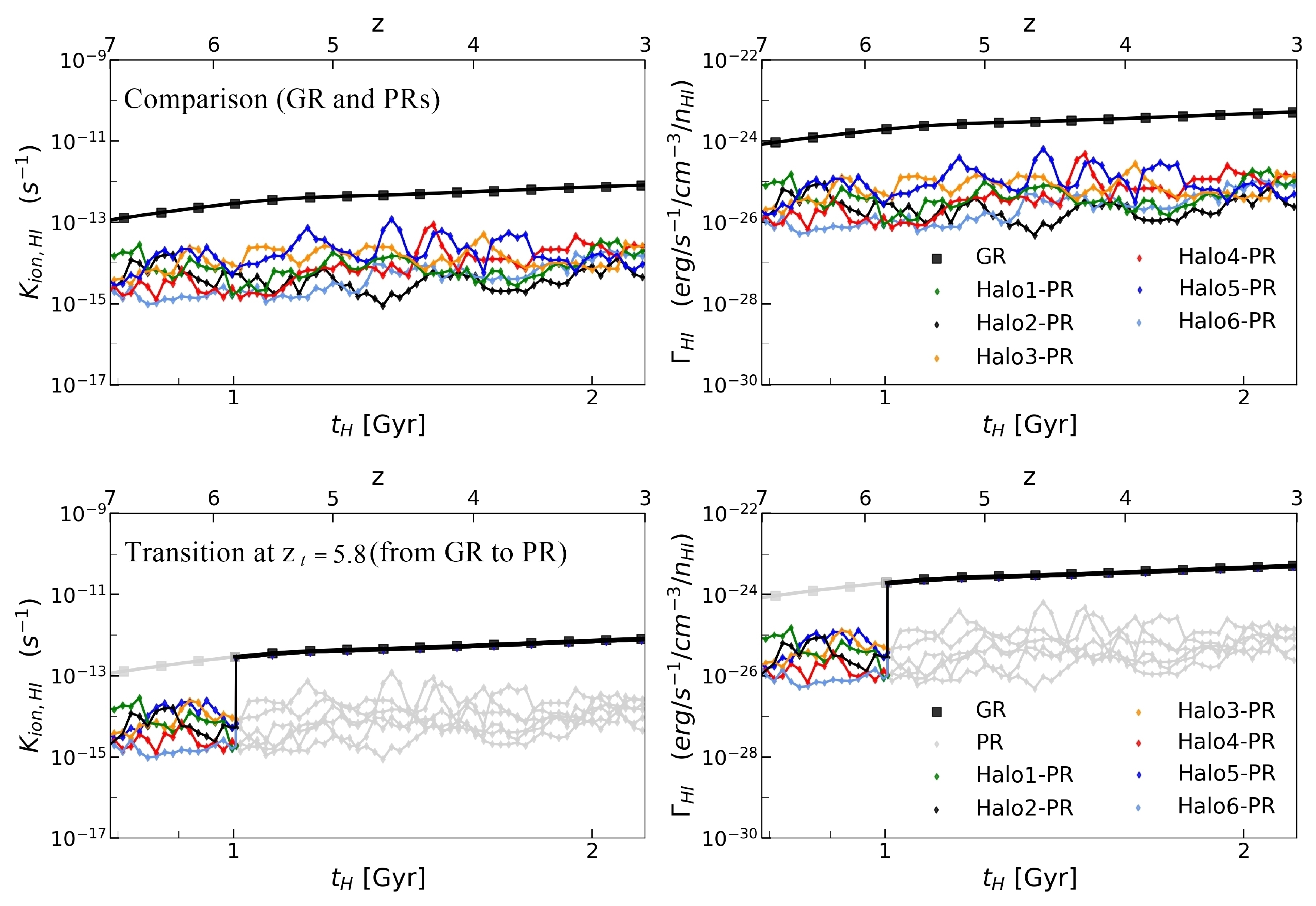}%
   \caption{The strength of patchy reionization in terms of hydrogen photoionization rate (left) and photoheating rate (right) exerted by a host galaxy onto a target UFD analog during the period from $z=7$ to $z=3$. The values derived from GR and PR runs are represented by solid black and colored lines, respectively. The PR values are, on average, two orders of magnitude lower than those of GR from the onset of the first star formation in UFD analogs ($z\gtrsim10$) until around $z\sim3$.} Since the contribution from star-forming galaxies to reionization is the only consideration, we switch to the GR implementation at a designated transition time, $z_{\rm t}=5.8$. In the bottom panels, the same data is presented as in the upper panels, but with a transition redshift of $z_{\rm t}=5.8$ applied. Below this redshift, the GR implementation is used, and this combination of GR and PR is referred to as PR going forward.
   \label{fig5}
\end{figure*}

Our analysis shows that the simulated galaxies in {\sc Halo2-GR}, {\sc Halo3-GR}, and {\sc Halo6-GR} have approximately three progenitors, while the stars in {\sc Halo1-GR}, {\sc Halo4-GR}, and {\sc Halo5-GR} are primarily ($\gtrsim75\%$) formed in a single primary halo. The difference between a multiple-progenitor and a single-progenitor group is attributed to the rate of halo mass growth. The primary halos of the single-progenitor group experience more rapid mass assembly than those of the multiple-progenitor group. For example, during the period of $z\sim10 - 7$, the primary halo of {\sc Halo4-GR} increases its halo mass from $M_{\rm vir}\sim4.8\times10^{6}\msun$ to $M_{\rm vir}\sim9.2\times10^{7}\msun$, along with an increase in gas mass, resulting in a substantial amount of star formation compared to the relatively less massive progenitor halos.

UFDs are known to be systems with low metallicity, with $\rm [Fe/H] \lesssim -2$ (e.g., \citealp{Martin2007, Norris2010, Kirby2013, Simon2019}). Due to the combined effects of reionization and SN feedback, which lead to short SFHs, the chances of these systems being enriched are limited. The metal-poor nature of UFD systems is depicted in Figure~\ref{fig4}. Within both the upper and bottom panels, the estimated $\rm [Fe/H]$ values of in-situ stars and externally originated stars are represented by black and grey inverted triangles, respectively, plotted as a function of their formation time. Note that in-situ stars refer to stars that are formed within the primary progenitor halo. We focus especially on two specific cases: in {\sc Halo2-GR}, stars form in multiple progenitors that merge at a later time, while in {\sc Halo4-GR}, stars are primarily formed in situ within the primary progenitor halo. In both halos, the $\rm [Fe/H]$ values of all stars show a wide range from $\rm [Fe/H]\approx-6$ to $\rm [Fe/H]\approx-2$ (upper panels). However, the metallicity of in-situ stars tends to display a narrow metallicity range of $\rm -3\lesssim [Fe/H] \lesssim -2$ (bottom panels). 

This trend is expected as star formation can proceed steadily in a relatively massive progenitor halo, increasing the metallicity of in-situ stars over time. This is because in the primary halos, the gas is not entirely expelled by the feedback of stars, and even if it is dispersed, it rapidly recollapses, allowing star formation to continue. Consequently, new stars can form with increased metallicity before the metals associated with the gas are diffused. Meanwhile, in relatively low-mass progenitor halos, star formation is more likely to be quenched by SN feedback, making it difficult to form stars with higher metallicities ($\rm [Fe/H]\gtrsim -2$). Furthermore, we observe that the formation of extremely low-metallicity stars ($\rm [Fe/H]\lesssim -5$) is solely a result of external metal enrichment. This occurs when gas is contaminated by metals from nearby halos, allowing for the formation of low-metallicity stars in the absence of previous metal-free star formation. Although the metallicity trend mentioned above is applicable to both simulations, {\sc Halo2-GR} and {\sc Halo4-GR}, the metallicity of {\sc Halo4-GR} (right panels) increases from [Fe/H] $\sim$ $-2.9$ to $-2.4$ and it shows a narrow range than {\sc Halo2-GR} due to the insufficient time for gas enrichment in the halo. This is because star formation begins later in {\sc Halo4-GR} than in {\sc Halo2-GR}. Specifically, {\sc Halo2-GR} starts forming the first Pop~II star at $z=13.55$, whereas in {\sc Halo4-GR}, the first Pop~II star is formed at $z=8.65$, around 270 Myr later. However, reionization at $z\sim7$ halts star formation in both simulations, resulting in a rather short metal enrichment history in {\sc Halo4-GR}.

\begin{figure*}
  \centering
  \includegraphics[width=130mm]{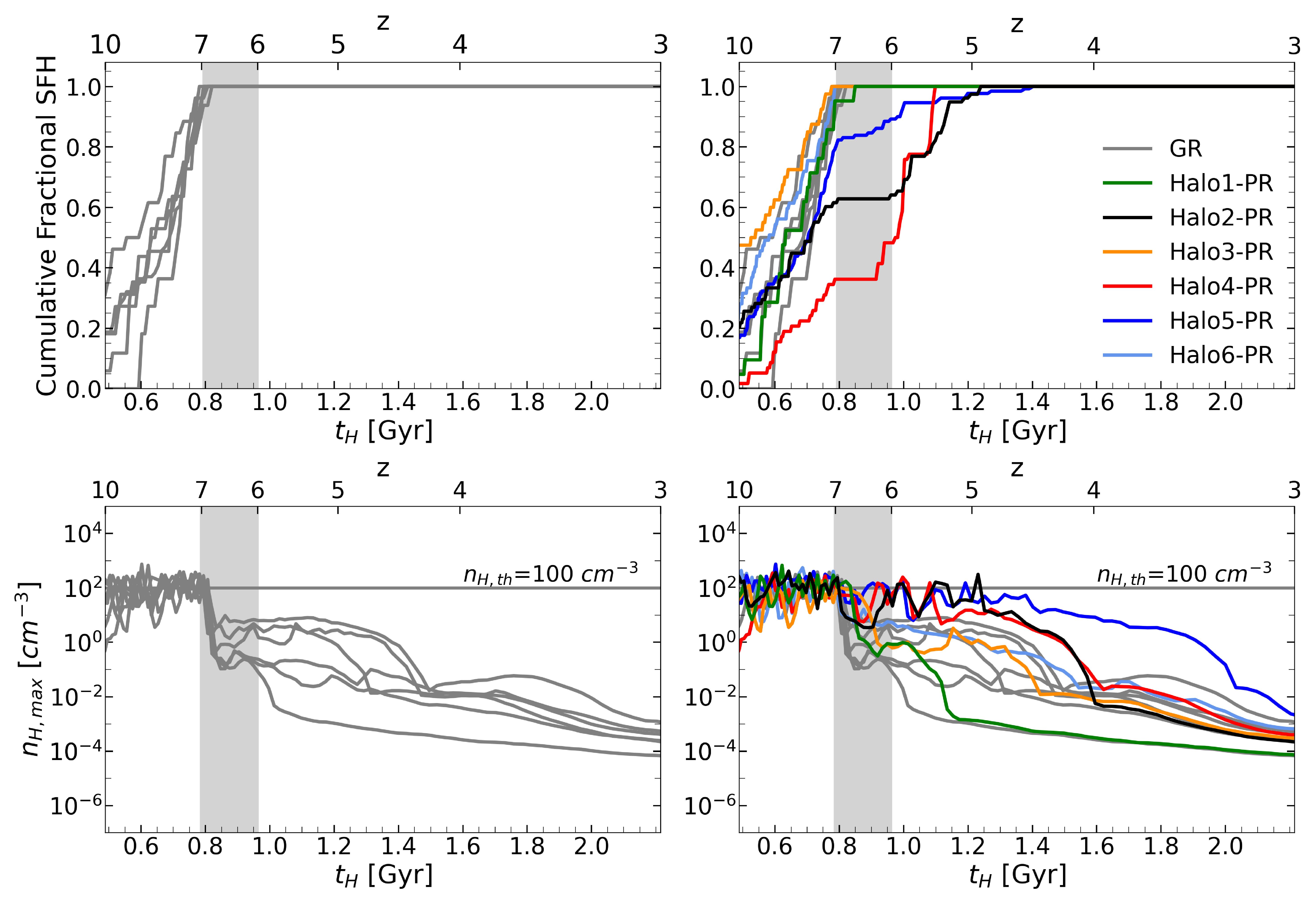}%
   \caption{Cumulative SFHs (top panel) and the maximum hydrogen number density of the gas (bottom panel) of the simulated UFD analogs, comparing the runs using GR (left panel) and PR (right panel) implementations. The evolution of each UFD analog is represented by different colors and line types. The simulations with GR implementation indicate complete cessation of star formation due to reionization, whereas three runs ({\sc Halo2-PR}, {\sc Halo4-PR}, and {\sc Halo5-PR}) using PR exhibit prolonged SFHs compared to those of GR runs. Among the runs adopting PR implementation, {\sc Halo5-PR} shows the most extended SFH, forming stars until around $z\sim4.45$. The late star formation phase due to patchy reionization lasts for 280 Myr, 180 Myr, and 550 Myr in {\sc Halo2-PR}, {\sc Halo4-PR}, and {\sc Halo5-PR}, respectively. The lower right panel indicates that halos with extended SFH tend to retain dense gas particles with a threshold density of $n_{\rm H, th}=100 \cmci$, which can facilitate late star formation due to the weaker UV field compared to that of GR.}
   \label{fig6}
\end{figure*}

\subsection{Patchy reionization effects on the SFHs and stellar metallicity} \label{sec:result_pr}

This section highlights the results of our galaxy simulations that are impacted by the spatially non-uniform reionization process. The intensity of this patchy reionization is ascertained by considering the environmental factors of each simulated galaxy, corresponding to the galaxy analogs listed in Table~\ref{table1}. As explained in Section~\ref{sec:Metho_uv}, we calculate the impact of patchy reionization by taking into account the stellar mass of the host halo and its distance as a function of cosmic time. The resulting photoionization and photoheating rates, covering the range from $z=7$ to $z=0$, are illustrated in Figure~\ref{fig5}. Although we consider the effects of PR on H~I, He~I, and He~II, only the results for H~I are presented in the upper panels of Figure~\ref{fig5} because we confirm that the effect of PR on the other species exhibits similar trends to that of H~I. The black solid line shows the values obtained from the GR runs, while the colored solid lines represent the estimates from the PR runs. The intensity of PR is, on average, two orders of magnitude lower than that of GR up to $z=3$, indicating that the effect of reionization on the simulated galaxies is relatively weaker when using the PR implementation. The calculated values should be regarded as the lower limit because we do not account for the complete photon flux originating from galaxies beyond the spatial scales we have modeled. Our findings indicate that the distant nature of these galaxies renders their impact insignificant, but the overall degree of reionization increases by a factor of $\sim5-10$ by considering galaxies on a larger scale (refer to Appendix~\ref{appendix_b} for details).

As described in Section~\ref{sec:Metho_uv}, ionizing sources contributing to reionization include radiation from both star-forming galaxies and quasars, with quasars being the predominant source at lower redshifts compared to star-forming galaxies. Given that we solely consider star-forming galaxies as the ionizing source, it is crucial to improve our PR estimates to account for the impact of quasars at low redshifts. As a result, we employ a hybrid method that merges GR and PR methods, wherein the reionization process transitions from PR to GR at a designated redshift. The lower panels of Figure~\ref{fig5} show the intensity of the combined GR+PR reionization. From this point onward, we will refer to GR+PR reionization as PR, and we have selected the transition time from PR to GR to occur at $z=5.8$. In the following two sections, we will explore the effects of PR on the SFH and stellar metallicity of the simulated galaxies.

\subsubsection{Star formation histories} \label{sec:result_pr_sf}

In Figure~\ref{fig6}, we present the resulting SFHs for both GR (left panels) and PR (right panels) implementations by showing the cumulative SFHs (top panels) and the maximum hydrogen number density (bottom panels) for our simulated halos. Furthermore, on the right panels, we overplot the values from the GR runs as grey lines to enable easy comparison. Similar to the GR runs, where all halos encounter a complete halt of star formation as a consequence of reionization, most halos except {\sc Halo5-PR} experience a temporary suppression of star formation with the patchy UV background. However, considering the reduced intensity of patchy reionization, which is two orders of magnitude weaker than that in the GR scenario, unlike the GR cases where the gas density fails to recover enough for star formation, in the PR cases, the gas density bounces back to a level that allows star formation to resume. We refer to this star formation taking place below $z=7$ in the PR runs as late star formation.

We observe that {\sc Halo2-PR}, {\sc Halo4-PR}, and {\sc Halo5-PR} form 60\%, 35\%, and 80\% of the total stellar mass before $z=7$ and undergo late star formation up to $z=4.45$ after reionization. Notably, {\sc Halo5-PR} exhibits a unique characteristic in its SFH, which extends 550 Myr since $z=7$, significantly longer than the 280 Myr in {\sc Halo2-PR} and 180 Myr in {\sc Halo4-PR} (see Table~\ref{table3}). Furthermore, in contrast to {\sc Halo2-PR} and {\sc Halo4-PR}, where late star formation commences at $z\sim6$ after a temporary quenching period from $z=7$ to $z=6$, {\sc Halo5-PR} manages to sustain star formation even when subjected to a stronger reionization effect than the other two cases, {\sc Halo2-PR} and {\sc Halo4-PR}.

As depicted in the bottom panels of Figure~\ref{fig6}, the maximum hydrogen number density drops to $n_{\rm H,max}=$ 0.1-1.0 $\rm cm^{-3}$ in the GR runs as soon as the reionization effect is introduced. On the other hand, in the PR runs, whether complete quenching occurs or late star formation is observed depends on the halo mass at the onset of reionization, corresponding to $z=7$. For instance, we find that three halos, {\sc Halo2-PR}, {\sc Halo4-PR}, and {\sc Halo5-PR}, with masses of $M_{\rm vir}\sim7.6\times10^{7}\msun$, $\sim9.7\times10^{7}\msun$, and $\sim1.6\times10^{8}\msun$ at $z=7$, respectively, exhibit late star formation. In contrast, relatively less massive halos, such as {\sc Halo1-PR} and {\sc Halo3-PR}, undergo complete quenching. Notably, {\sc Halo5-PR} is the most massive halo at $z=7$ among the three halos ({\sc Halo2-PR}, {\sc Halo4-PR}, and {\sc Halo5-PR}). By maintaining high-density gas particles within its virial radius, {\sc Halo5-PR} can sustain continuous star formation.

We find that stars produced during the late star formation period contribute to 40\% ({\sc Halo2-PR}), 65\% ({\sc Halo4-PR}), and 20\% ({\sc Halo5-PR}) of the total stellar mass in each respective run. Interestingly, there seems to be a negative correlation between the duration of late star formation and the fraction of stars formed during that period, with {\sc Halo5-PR} showing the lowest value of the fraction of late star formation; however, this correlation is not meaningful. The fraction of stars formed through late star formation within a given halo is determined by a complex interplay of factors, including the halo mass at the onset of reionization, the total duration of star formation activity, and the burstiness of star formation. It is important to note that the galaxies in {\sc Halo2-PR}, {\sc Halo4-PR}, and {\sc Halo5-PR} attain comparable halo masses at $z=5$. Nonetheless, the fraction of stars formed in late star formation can differ significantly based on when the halo growth occurs, either before or after reionization. For example, {\sc Halo2-PR} and {\sc Halo4-PR} form a substantial amount of stars after reionization, whereas in the case of {\sc Halo5-PR}, as mentioned earlier, even though the galaxy exhibits the most extended SFH until $z\sim4.45$, the majority of its stars ($\sim80\%$) have already formed prior to reionization.

Furthermore, while {\sc Halo4-PR} has a shorter duration of late star formation of $\sim$180 Myr, around 100 Myr less than that of {\sc Halo2-PR}, a larger fraction of stars are formed in the late phase in {\sc Halo4-PR} (roughly 65\%) compared to {\sc Halo2-PR}, which forms about 40\% of its stars below $z=7$. This difference can be ascribed to the galaxy in {\sc Halo4-PR} commencing star formation at a later redshift ($z\approx8$) relative to the galaxy in {\sc Halo2-PR}. Moreover, the galaxy in {\sc Halo4-PR} undergoes a phase of bursty star formation at $z=6$, during which stars with a total mass of $M_{\ast}\approx1.1\times10^4\msun$ form within a 10 Myr span. Consequently, the duration of late star formation in {\sc Halo4-PR} is shorter than that of {\sc Halo2-PR}, due to the strong stellar feedback from the bursty star formation event that interrupts subsequent star formation.

It should be noted that using the PR approach in simulating UFD analogs does not guarantee extended SFHs. Achieving late star formation becomes challenging if the progenitor halo's mass is relatively small, leading to complete quenching despite the weak intensity from PR. For example, as illustrated in Figure~\ref{fig5}, the PR intensity applied to the galaxy in {\sc Halo3-PR} is comparable to that of {\sc Halo5-PR} between $z=7$ and $z=6$. However, the halo mass of {\sc Halo3-PR} ($M_{\rm vir}\sim7.0\times10^{7}\msun$) is 2.3 times less massive than that of {\sc Halo5-PR} at $z=7$, which makes it difficult for the galaxy to maintain gas particles able to form stars after reionization. Consequently, the mass of the progenitor halo at the time of onset of reionization is critical for an extended SFH.

\begin{figure}
  \centering
  \includegraphics[width=80mm]{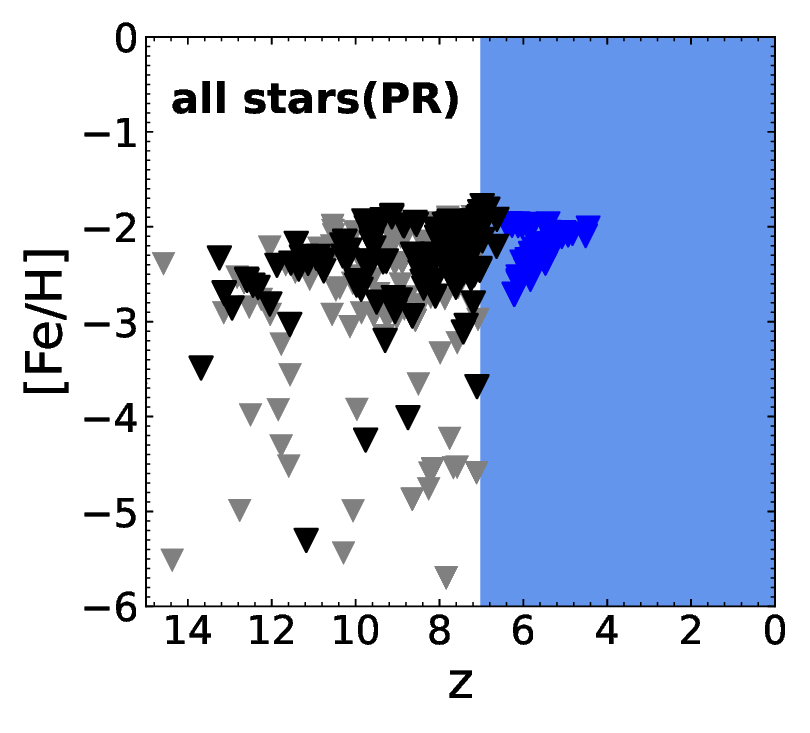}%
   \caption{The same as Figure~\ref{fig4}, but it displays the metallicity for stars from all PR runs. The blue-shaded region in each panel indicates the period where the patchy reionization effect is applied, and the blue inverted triangles within this region represent the stars formed during the late star formation phase in all PR runs. It is interesting to note that the stars formed during the late phase of star formation are likely to exhibit metallicity levels similar to or even lower than those of stars with the highest metallicity formed prior to reionization.} This is because even though the UV field from patchy reionization is not strong enough to completely stop star formation, it temporarily halts it between $z=7$ and $z=6$. During this time, metals in the gas tend to diffuse along with the gas due to patchy reionization. This leads to a decrease in the gas metallicity and the formation of stars with similar or lower metallicities.
   \label{fig7}
\end{figure}

\begin{figure*}
  \centering
  \includegraphics[width=135mm]{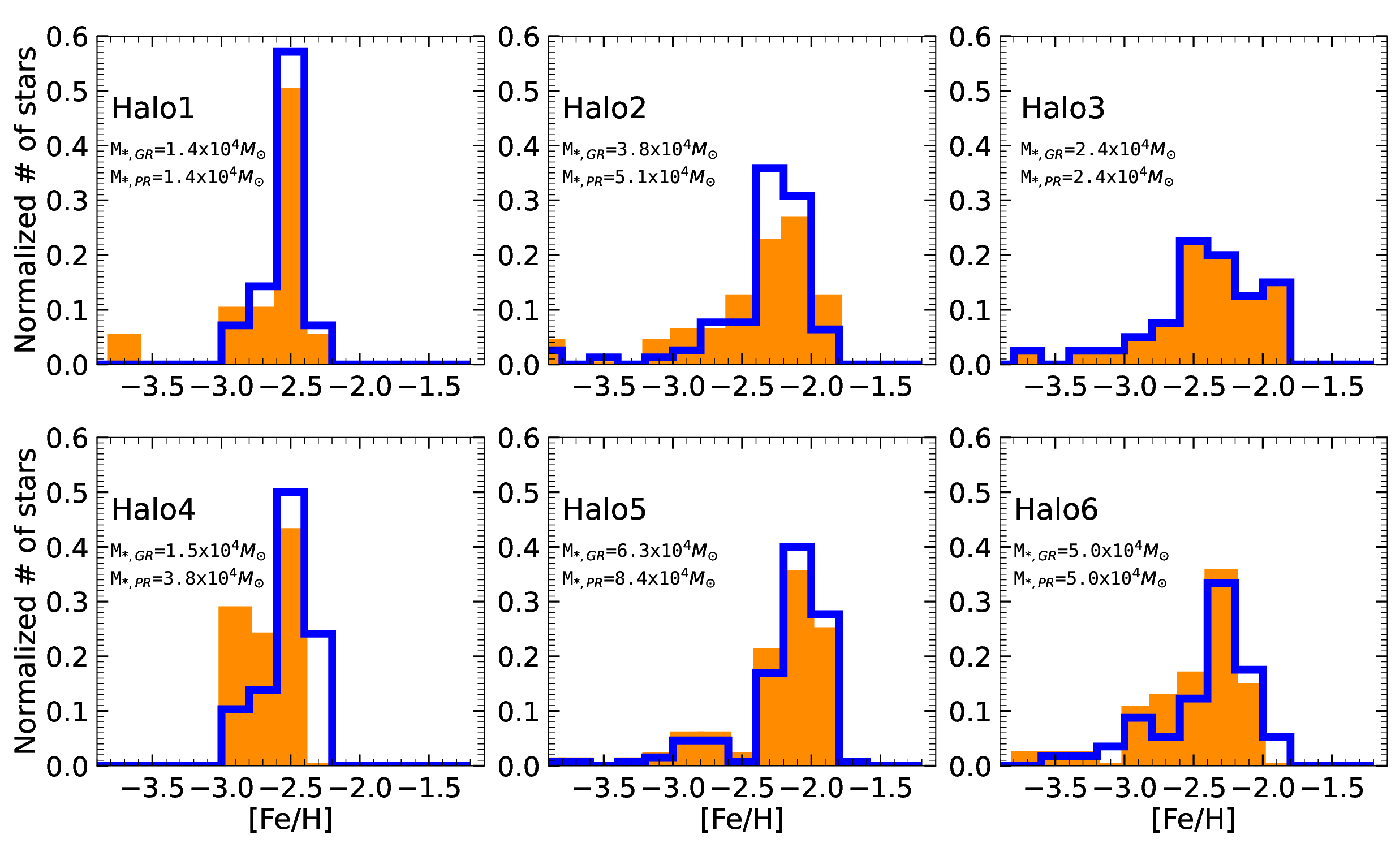}%
   \caption{The normalized metallicity distribution function of stars is shown for each simulated galaxy, with the range spanning from $\rm [Fe/H]=-3.9$ to $\rm [Fe/H]=-1.1$ and an interval of 0.2 dex. The orange (blue) histogram exhibits the simulated halo adopting the GR (PR) implementation. In each halo, the peak of the MDF of stars ranges from $\rm [Fe/H]=-2.5$ to $\rm [Fe/H]=-2$. The MDFs of halos with PR implementation, especially {\sc Halo2-PR, Halo4-PR, Halo5-PR}, display a shift towards higher metallicity compared to those with GR. This shift is attributed to the extended star formation period in the PR runs, which leads to the formation of stars with higher metallicity.}
   \label{fig8}
\end{figure*}

\subsubsection{Stellar metallicity} \label{sec:result_pr_metal}
To investigate whether the prolonged SFHs resulting from weaker PR intensities are reflected in the stellar metallicity of the simulated galaxies, we present Figure~\ref{fig7}. This figure illustrates the stellar metallicity as a function of the formation time of stars located within the virial radius of the simulated halos using PR at $z=0$. In particular, we compare the [Fe/H] values for all stars (gray and black) within the virial radius of all the simulated galaxies employing PR to those formed internally within the primary progenitor halos (black). We find that almost all stars are metal-poor, ranging from $\rm [Fe/H] = -2$ to $\rm [Fe/H] = -5$, and the in-situ stars (black) also exhibit a similar metallicity range to that of all stars. We denote the stars formed during late star formation with blue inverted triangles in each panel. The metallicity of these late-forming stars, ranging between $\rm -3\lesssim[Fe/H]\lesssim-2$, is comparable to that of in-situ stars formed at $z\sim7$, just before a patchy UV field is introduced.

We observe two distinct characteristics of stars resulting from late star formation. Firstly, all stars originating from the late star formation phase are formed internally within the halo rather than being accreted stars. This occurs because only the primary halo can sustain star formation under the influence of reionization, while less massive progenitors experience a total cessation of star formation due to reionization. Secondly, even though these newly formed in-situ stars originate at relatively lower redshifts, they tend to display metallicities similar to or lower than the peak [Fe/H] of stars formed prior to reionization. This is attributed to the temporary quenching experienced by the simulated galaxies, even with the weak impact of patchy reionization, which causes dense gas and metals to dissipate. Given that the late star formation in {\sc Halo2-PR} and {\sc Halo4-PR} commences later than in {\sc Halo5-PR}, in-situ stars from the late star formation at $z\sim6$ in {\sc Halo2-PR} and {\sc Halo4-PR} exhibit lower metallicities ([Fe/H] $\sim$ -2.7) than those in {\sc Halo5-PR} by approximately 0.6 dex. This implies that the longer it takes for the late star formation to begin, the more dense gas in gas in the primary progenitor halo disperses. Consequently, it becomes difficult to form high metallicity stars ($\rm [Fe/H]\sim-2$). Once the late star formation starts, the peak [Fe/H] of in-situ stars increases from $\rm [Fe/H]=-2.7$ to $\rm [Fe/H]=-2.2$ over time in both {\sc Halo2-PR} and {\sc Halo4-PR} runs, due to the increase in metals from SN explosions occurring within the halo.

\begin{figure*}
  \centering
  \includegraphics[width=140mm]{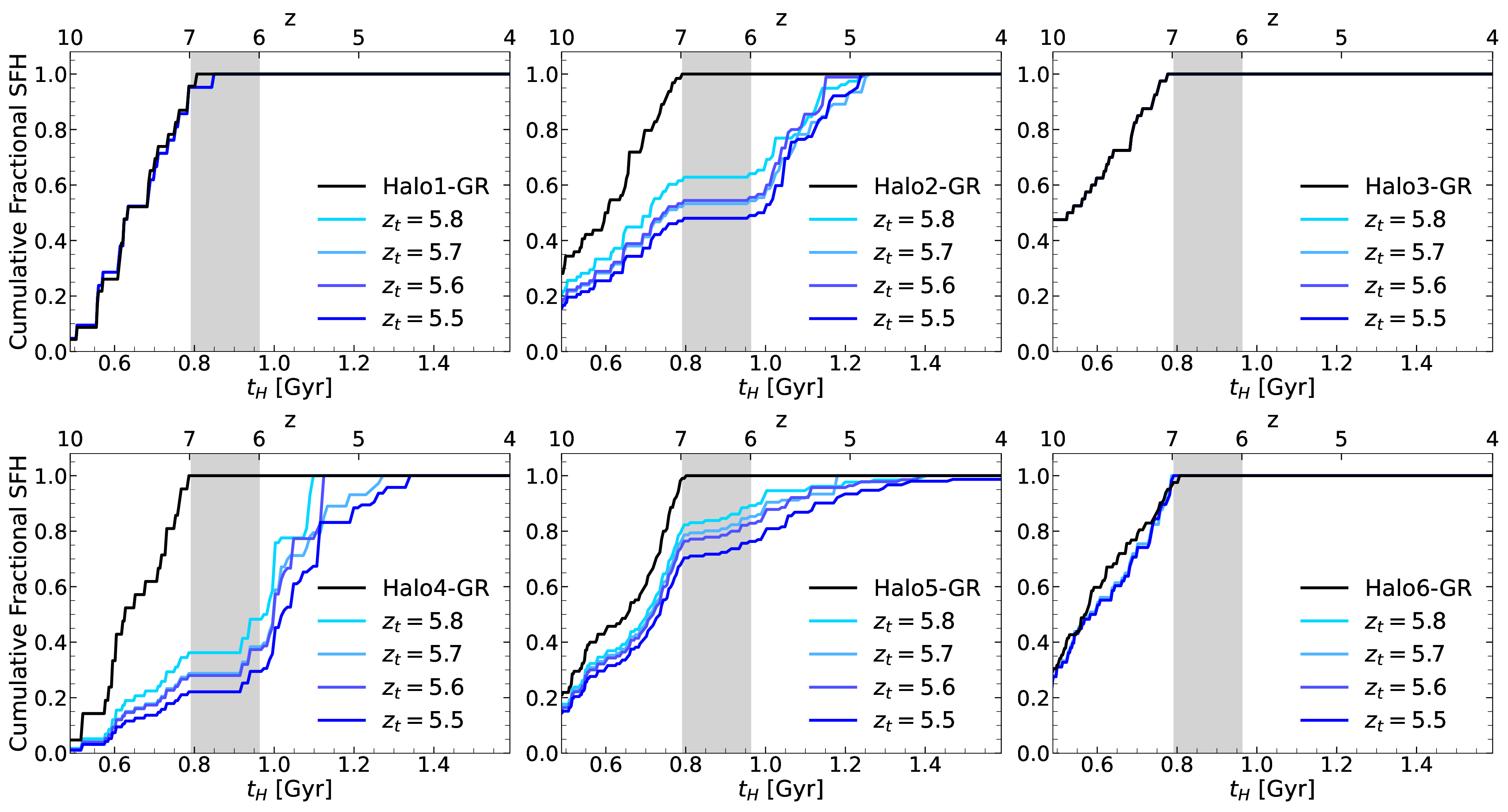}%
   \caption{ The cumulative SFHs of the simulated galaxies are shown as the transition time decreases from $z_{\rm t}=5.8$ to $z_{\rm t}=5.5$. As the completion of reionization is delayed, the halos tend to form more stars through late star formation, and the duration of the late star formation generally increases. Consequently, the stars produced by late star formation contribute to an increase in stellar mass. For example, when the transition time is changed from $z_{\rm t}=5.8$ to $z_{\rm t}=5.5$, the stellar mass of Halo2-PR and Halo4-PR increases by up to 31\% and 63\%, respectively. Furthermore, our findings suggest that the extent of late star formation in the simulated galaxies is not only dependent on the transition time from PR to GR implementation but also on the nature of the late star formation, whether it occurs in episodic bursts or as a continuous process.}
   \label{fig9}
\end{figure*}

In Figure~\ref{fig8}, we show the metallicity distribution function (MDF) of the simulated galaxies, where we categorize Pop II stars from each halo, spanning from $\rm [Fe/H]=-3.9$ to $\rm [Fe/H]=-1.1$, with intervals of 0.2 dex. Specifically, we compare the MDF of Pop II stars from GR runs, depicted in orange, with those from PR runs, illustrated in blue. First, the MDF of galaxies with a stellar mass larger than $M_{\rm *} > 10^{4}\msun$ at $z=0$ tends to peak at $\rm [Fe/H]\sim-2$. We observe no significant difference in the MDF, except for {\sc Halo4-PR}, among the three runs: {\sc Halo2-PR}, {\sc Halo4-PR}, and {\sc Halo5-PR}, which have extended SFHs due to adopting patchy reionization. Only in the run {\sc Halo4-PR} do we find the MDF shifts as a result of the patchy reionization effect, extending towards a higher metallicity by 0.22 dex. The cause of this minimal change in the MDF, shown in {\sc Halo2-PR} and {\sc Halo5-PR}, is that the simulated galaxies experience a temporary quenching due to weak patchy reionization rather than continuously forming stars. Consequently, metals are expelled along with the gas, reducing the gas metallicity when the halo is replenished to form subsequent stars.

\subsection{Patchy UV background with transition time} \label{sec:result_zt}
In this section, the question we aim to address is how the duration of late star formation and stellar metallicity may change if we adopt a transition time for the PR to GR implementation up to $z_{\rm t}=5.5$, which is later than the fiducial value of $z_{\rm t}=5.8$. To investigate this, we have conducted simulations on the same runs as in the PR cases, but with a gradual change in the transition time from $z_{\rm t}=5.8$ to $z_{\rm t}=5.5$, incrementing by 0.1. We explore the effects of a delayed transition time for two primary reasons: (1) there is no consensus on when reionization is completed, and (2) recent observational studies have proposed a late reionization scenario, indicating that reionization may have been completed between $z=5.5$ and $z=6$ (e.g., \citealp{Becker2015}; \citealp{Choudhury2015}; \citealp{McGreer2015}; \citealp{Mesinger2015}). As such, by incorporating a delayed transition time, we are also able to examine the implications of a late reionization scenario. The properties of the simulated halos with delayed transition times are summarized in Table~\ref{table3}. As explained in Section~\ref{sec:result_pr}, the notation PR-z`5.x' in the halo name signifies that the transition from PR to GR takes place at $z=5$.x. For example, in the PR-z5.5 run, the transition from PR to GR occurs 0.3 later in redshift compared to the PR-z5.8 run, resulting in a longer application of the PR effect. It is important to note that all physical quantities in Table~\ref{table3} are calculated based on the particles within the virial radius of each halo at $z=3$. We halted all simulations adopting PR at $z=3$ because no more late star formation is expected in the simulated halos beyond this period, as demonstrated in the GR runs.

Figure~\ref{fig9} presents the SFHs of the simulated halos as a function of the transition time. Notably, we find that in line with the fiducial PR runs, the relatively less massive halos during the onset of reionization ({\sc Halo1-PR, Halo3-PR, Halo6-PR}) still fail to achieve sufficient density for star formation to take place, even when the patchy reionization period is prolonged by postponing the transition time. Meanwhile, the impact of the late transition time is evident in the runs {\sc Halo2-PR, Halo4-PR}, and {\sc Halo5-PR} as follows. As expected, when the period of patchy reionization is extended by delaying the transition from PR to GR, the simulated galaxies tend to form more stars through late star formation.

For example, with the transition time of $z_{\rm t}=5.8$ ($z_{\rm t}=5.5$), {\sc Halo2-PR} in Figure~\ref{fig9} accounts for 40\% (50\%) of stars from late star formation, so more stars are formed in {\sc Halo2} by late reionization. Consequently, the stars produced by late star formation contribute to an increase in stellar mass. For instance, when adopting ($z_{\rm t}=5.5$), the stellar mass of {\sc Halo2-PR} increases by up to 31\%, comapared to those of {\sc Halo2-PR-z5.8} (refer to Table~\ref{table3}).

We observe a slight correlation between the transition time and the duration of late star formation. As indicated in Table~\ref{table3}, by postponing the transition time from $z_{\rm t}=5.8$ to $z_{\rm t}=5.5$, the duration of late star formation in {\sc Halo4-PR} can be extended by approximately 240 Myr, while {\sc Halo2-PR} experiences a minimal change in duration, at around 15 Myr. Interestingly, the duration of late star formation and the resulting stellar masses are not always proportional, as evidenced by the weak correlation in {\sc Halo4-PR}. To be specific, the stellar mass of {\sc Halo4-PR-z5.6} is similar to that of {\sc Halo4-PR-z5.7}, with a difference in halo mass of less than 3\%. However, {\sc Halo4-PR-z5.6} has a late star formation duration that is shorter than {\sc Halo4-PR-z5.7} by approximately 140 Myr. 

This difference is ascribed to whether the galaxy undergoes bursty or continuous star formation. In {\sc Halo4-PR-z5.6}, stars with a total stellar mass of $M_{\ast}=1.2\times10^4\msun$ are formed during a short period of 4.4 Myr, causing the galaxy to experience a more episodic and bursty star formation compared to {\sc Halo4-PR-z5.7}. Stellar feedback from such bursty star formation significantly reduces the gas density by two orders of magnitude, effectively suppressing further star formation. Overall, the extent of the late star formation of the simulated galaxies depends not only on the transition time from PR to GR implementation but also on the nature of their late star formation, whether occurring in episodic bursts or as a continuous process. To examine the primary factor shaping the characteristics of SFHs, we carry out additional simulations employing various star formation random seeds. Our results reveal that the trends, which suggest that a delayed transition results in a more prolonged SFH, remain consistent, albeit with slight variations in the duration of late star formation due to its inherent nature.

\begin{figure}
  \centering
  \includegraphics[width=70mm]{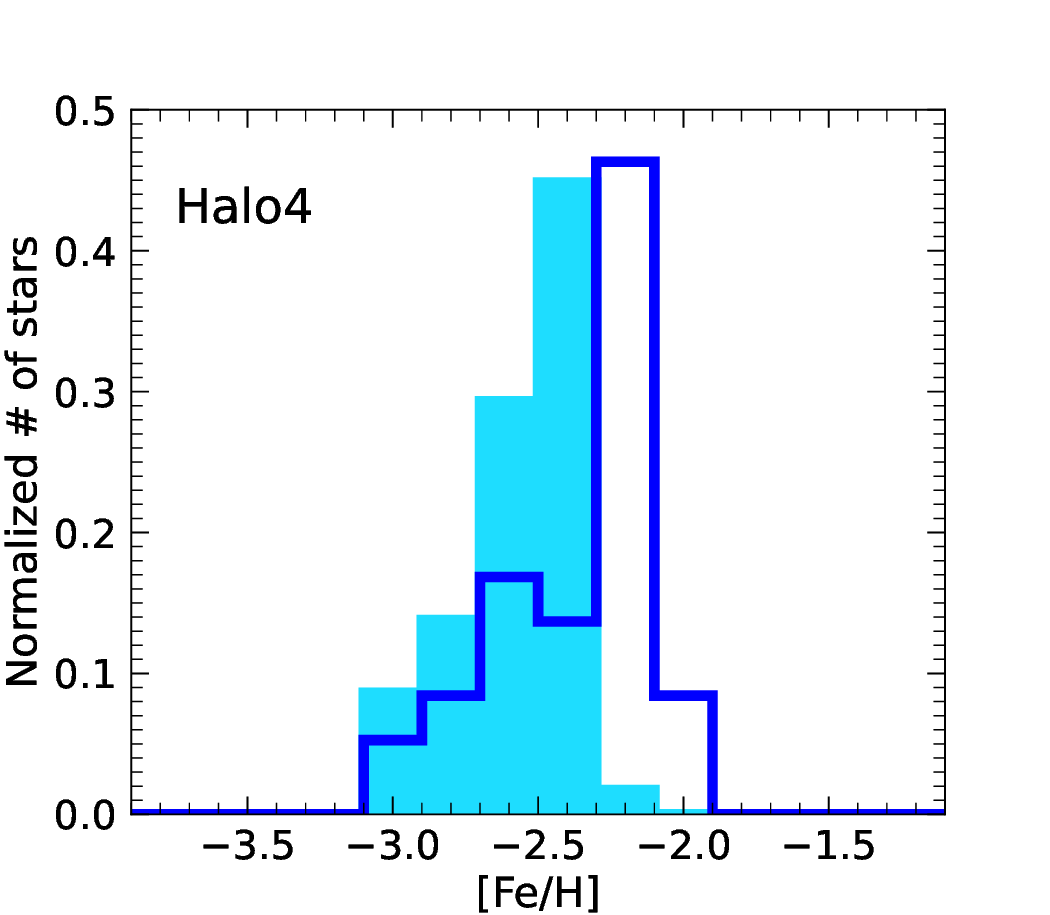}%
   \caption{The normalized MDF of stars from {\sc Halo4-PR} is shown to compare the effect of transition time, with the cyan-filled MDF corresponding to $z_{\rm t}=5.8$, and the blue MDF representing the run adopting $z_{\rm t}=5.5$. As the transition time is moved to lower redshift, the fraction of late star formation in {\sc Halo4-PR} increases from 65\% to 80\%. This contributes to the shift towards higher metallicity in the MDF, as previously discussed in Figure~\ref{fig8}. For {\sc Halo4-PR}, we have found that delaying the transition time from $z_{\rm t}=5.8$ to $z_{\rm t}=5.5$ results in a mean stellar metallicity increase of 0.2 dex.}
   \label{fig10}
\end{figure}

Since {\sc Halo2-PR} and {\sc Halo4-PR} have similar masses at the onset of reionization, it is useful to compare their evolution while keeping the transition time constant, given that the intensity of patchy reionization in {\sc Halo4-PR} is one-third that of {\sc Halo2-PR}. Our comparison reveals that when $z_{\rm t}=5.5$ is employed, the SFH of {\sc Halo4-PR} extends for 90 Myrs longer than that of {\sc Halo2-PR}. Furthermore, it would be meaningful to consider {\sc Halo2-PR} and {\sc Halo4-PR} as non-Magellanic and Magellanic UFDs, respectively, as proposed by \citet{Sachhi2021}. Assuming that Magellanic UFDs may have been situated farther from the host halo during reionization, the strength of reionization could be weaker than that applied to non-Magellanic satellites that had already entered the host environment at that time. Consequently, Magellanic systems could be expected to exhibit longer SFHs, similar to {\sc Halo4-PR}.

To explore how the stellar metallicity may vary based on the transition time, we compare the MDF of Pop~II stars in {\sc Halo4-PR-z5.8} (filled cyan) and {\sc Halo4-PR-z5.5} (blue) in Figure~\ref{fig10}. Notably, it is evident that the relatively high metallicity stars with $\rm [Fe/H]\gtrsim-2.3$ are formed only in the {\sc Halo4-PR-z5.5} run during late star formation, resulting in a median metallicity of $\rm [Fe/H]=-2.24$ which is greater than the median value of $\rm [Fe/H]=-2.51$ shown in {\sc Halo4-PR-z5.8}. As discussed in Section~\ref{sec:result_pr_metal}, if the duration of late star formation resulting from weak patchy reionization is temporary, it may be difficult to discern a significant difference in the subsequent stellar metallicity. This is because metals are expelled alongside the gas during the short quenching period, leading to the formation of stars with metallicities that are either lower or similar to those formed prior to reionization. On the other hand, if the duration of late star formation is prolonged by postponing the transition time, the patchy reionization effect becomes more apparent in the emerging stellar metallicity.

\subsection{Comparison with observations and theoretical work} \label{sec:result_obs}
So far, we have demonstrated the impact of patchy reionization on the SFHs and stellar metallicity of UFD galaxy analogs while taking into account environmental conditions. In this section, we will compare our simulated UFD analogs, which incorporate a transition redshift of $z_{\rm t}=5.5$, with observed UFD satellites in the MW. Additionally, we will compare our findings with other theoretical studies that have investigated the metallicity of UFDs. The objective is to provide novel insights into star formation, stellar metallicity, and the galactocentric distance of observed UFD satellites through the implementation of our patchy reionization model.

\begin{figure}
  \centering
  \includegraphics[width=80mm]{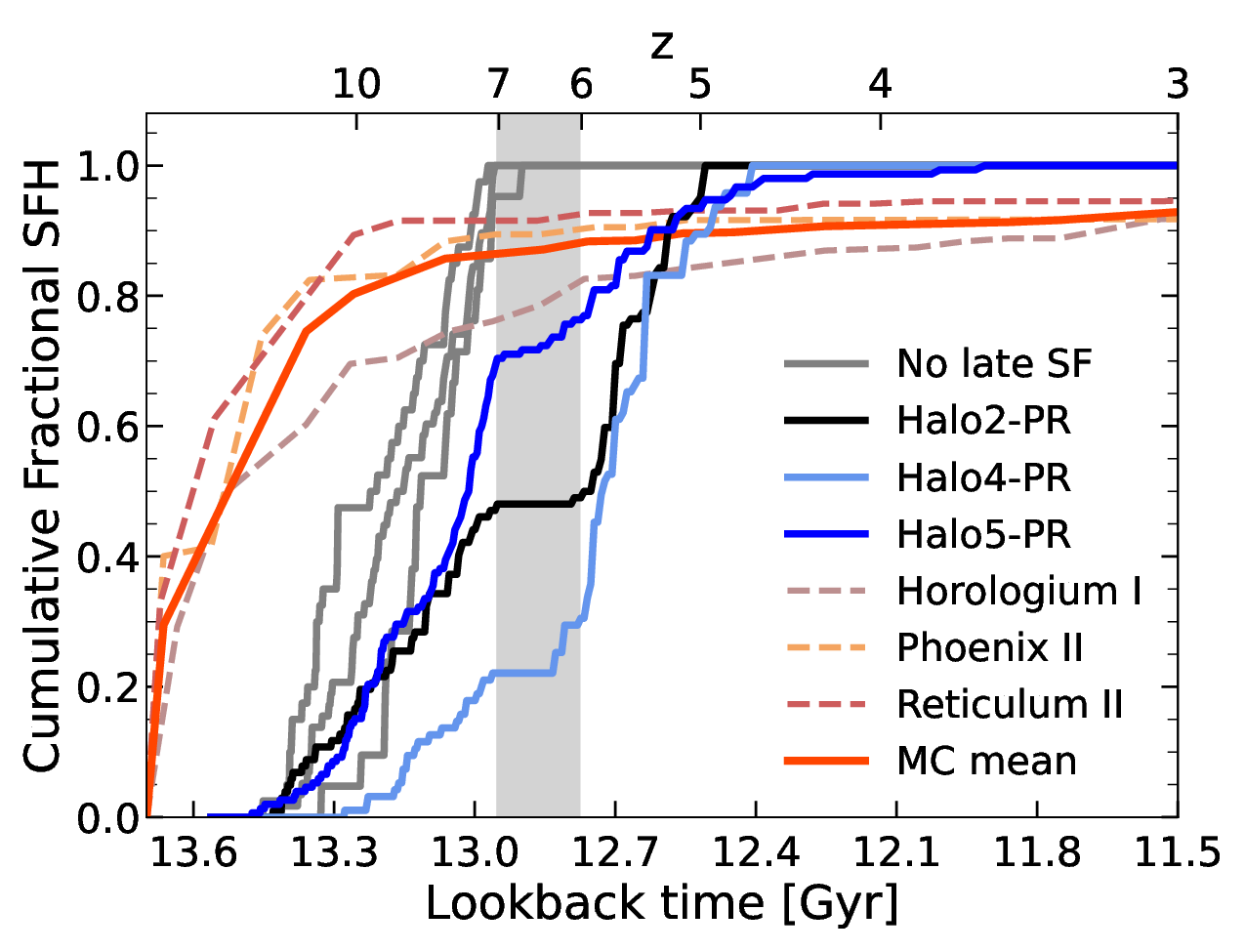}%
   \caption{Cumulative fractional SFHs of the simulated UFD analogs, alongside those of observed MW UFD satellites, represented by different colors. The PR runs are distinguished by colored lines if they exhibit extended SFHs, while gray is used for PR runs without extended SFHs. Here, we focus on the cumulative SFHs of Magellanic UFDs, which are believed to have recently entered the MW environment and are, therefore, suitable samples for studying the effects of patchy reionization. Compared to our simulations, the observed UFDs tend to have formed a majority of their stars ($\sim90\%$) prior to reionization, but they also exhibit extended star formation that continues for at least 3 Gyr longer than what is found in our simulated halos.}
   \label{fig11}
\end{figure}

\begin{deluxetable}{cccc}
\tablecaption{The quenching times and the completion time of star formation in Ultra-Faint Dwarf galaxies (UFDs) are provided on average\label{table2}}
\tabletypesize{\footnotesize}
\tablewidth{0pt}
\tablehead{
\colhead{Galaxy} & \colhead{$\tau_{\rm 50}$(Gyr ago)} & \colhead{$\tau_{\rm 90}$(Gyr ago)} & \colhead{$\rm SF_{\rm end}$(Gyr ago)}
}
\startdata
non-MC mean  & 13.40 $\pm$ 0.06 & 12.68 $\pm$ 0.23  &  11.5 \\
MC mean  & 13.49 $\pm$ 0.09 & 12.06 $\pm$ 0.72 & 8.5 \\
\hline
GR mean & 13.15 & 13.01 &  12.96  \\
PR-z5.5 mean  & 12.82 & 12.55 &  12.28  \\
\enddata
\tablecomments{Column (1) specifies the names of each UFD group. Columns (2), (3), and (4) represent the times at which UFDs reach 50\%, 90\%, and 100\% of their final stellar mass, respectively.}
\end{deluxetable}

\subsubsection{Star formation histories of the observed UFDs} \label{sec:result_obs_sfh}
Figure~\ref{fig11} depicts a comparison of the cumulative fractional SFH obtained from simulated galaxies with those of Magellanic UFDs, including Horologium~I, Phoenix~II, and Reticulum~II, along with their averaged SFH denoted as MC mean (\citealp{Sachhi2021}). We have specifically selected the simulation runs that incorporate patchy reionization with a transition time of $z_{\rm t}=5.5$, as these runs demonstrate the most extended SFHs. Based on the reconstructed SFHs of observed Magellanic UFDs, it is likely that they formed a significant proportion of stars (mean value of $\sim80\%$) before $z=7$, while the simulated galaxies tend to display a broader range in terms of the fraction of stars formed prior to $z=7$, which spans from 20\% to 70\%. In addition, the observed Magellanic UFDs are expected to continue forming the remaining 20\% of stars for a more extended duration, up to a redshift of $z=1.2$, while the simulated galaxies complete their star formation by $z\approx3.56$.

Table~\ref{table2} shows a summary of the differences between the simulated galaxies and observed Magellanic UFDs in terms of the quenching time, including $\tau_{\rm 50}$ and $\tau_{\rm 90}$, which represent the time required to form 50\% and 90\% of the final stellar mass (e.g., \citealp{Weisz2019}), and $\rm SF_{\rm end}$, which indicates the point at which star formation is entirely quenched, both as a look-back cosmic time. The table reveals that the quenching time, $\tau_{\rm 90}$, is similar for Magellanic UFDs and {\sc PR-z5.5} runs with $\tau_{\rm 90}=12.68$ Gyr ago and $\tau_{\rm 90}=12.55$ Gyr ago, respectively, implying that 90\% of stars are formed by $z\approx5$ in both cases. However, the timing of the most rapid star formation is different between the two, with Magellanic UFDs forming stars early, represented as $\tau_{\rm 50}=13.5$ Gyr ago, while it occurs relatively late, showing $\tau_{\rm 50}=12.7$ Gyr ago for the {\sc PR-z5.5} runs. As discussed in Section~\ref{sec:result_pr_sf}, the fraction of stars formed before cosmic reionization is greater when the halo mass is higher, particularly in the early stages of its evolution, which is evident in the case of {\sc Halo~5}. Conversely, for smaller halo masses during the early stages of evolution, such as {\sc Halo~2} and {\sc Halo~4}, the impact of patchy reionization is more significant. The implication is that patchy reionization has the potential to enable galaxies that would have otherwise experienced complete quenching $z \sim 7$ with the GR model, to form more stars instead. This, in turn, leads to a smaller value of $\tau_{\rm 50}$.

Magellanic UFDs exhibit a more prolonged star formation period compared to the PR runs. For instance, Magellanic UFDs with $\rm SF_{\rm end}=8.5$ Gyr ago have SFHs that are extended by at least 3 Gyr compared to those of the simulated galaxies. The mean value for $\rm SF_{\rm end}$ of the PR runs with $z_{\rm t}=5.5$ is $\rm SF_{\rm end}=12.5$ Gyr ago. It is possible that the difference in $\rm SF_{\rm end}$ between Magellanic UFDs and the simulated galaxies could be due to weaker reionization experienced by the former from the surrounding galaxies. This suggests that the strength of patchy reionization applied in our {\sc PR-z5.5} runs might be greater than what was experienced by the Magellanic UFDs. Furthermore, as discussed in Section~\ref{sec:result_pr_sf}, comparing {\sc Halo2} and {\sc Halo4}, the duration of star formation can also depend on whether the star formation is continuous or bursty, while patchy reionization is taking place even for halos of the same mass. In other words, the more bursty the star formation, the greater the impact of the powerful SN explosion effect, resulting in complete quenching and, consequently, a shorter duration of star formation.

Although our simulations may not perfectly replicate the extent of star formation in the observed Magellanic UFDs, the fact that patchy reionization can extend the duration of star formation to a greater extent than homogeneous reionization is in line with the findings of \citet{Sachhi2021}. For instance, the quenching times of the simulated UFDs with extended SFHs ({\sc Halo2-PR}, {\sc Halo4-PR}, and {\sc Halo5-PR}), on average, are $\tau_{\rm 90}=12.55$ Gyr ago, compared to $\tau_{\rm 90}=13.01$ Gyr ago for the runs with homogeneous reionization, resulting in a difference of 460 Myr. The observed difference in quenching time between non-Magellanic and Magellanic UFDs, approximately 600 Myr, aligns with our findings and lends support to theoretical models suggesting that the SFHs of MW satellite galaxies may exhibit hints of patchy reionization during the early Universe.

In summary, our simulation results suggest that the extended SFHs of Magellanic UFDs, as reconstructed through a color-magnitude diagram, may be attributed to a non-uniform reionization effect or halo mass. Galaxies within massive halos can overcome the suppression of star formation by reionization, leading to longer star formation duration. However, given the negligible difference in stellar masses between Magellanic and non-Magellanic UFDs, the prolonged SFHs of Magellanic UFDs are more likely due to patchy reionization effects. The fact that approximately 90\% of the stellar mass of Magellanic UFDs is estimated to have formed before reionization, and the remaining 10\% during the period of patchy reionization, suggests that the intensity of patchy reionization was moderate, and star formation was not bursty. If star formation were to burst, the dense gas would have been dissipated by the powerful SN feedback, leading to shortened SFHs. Alternatively, it could be interpreted that the transition from patchy to global reionization occurred at a late time. However, such a scenario would result in higher stellar masses than those observed in Magellanic UFDs.

\begin{figure}
  \centering
  \includegraphics[width=80mm]{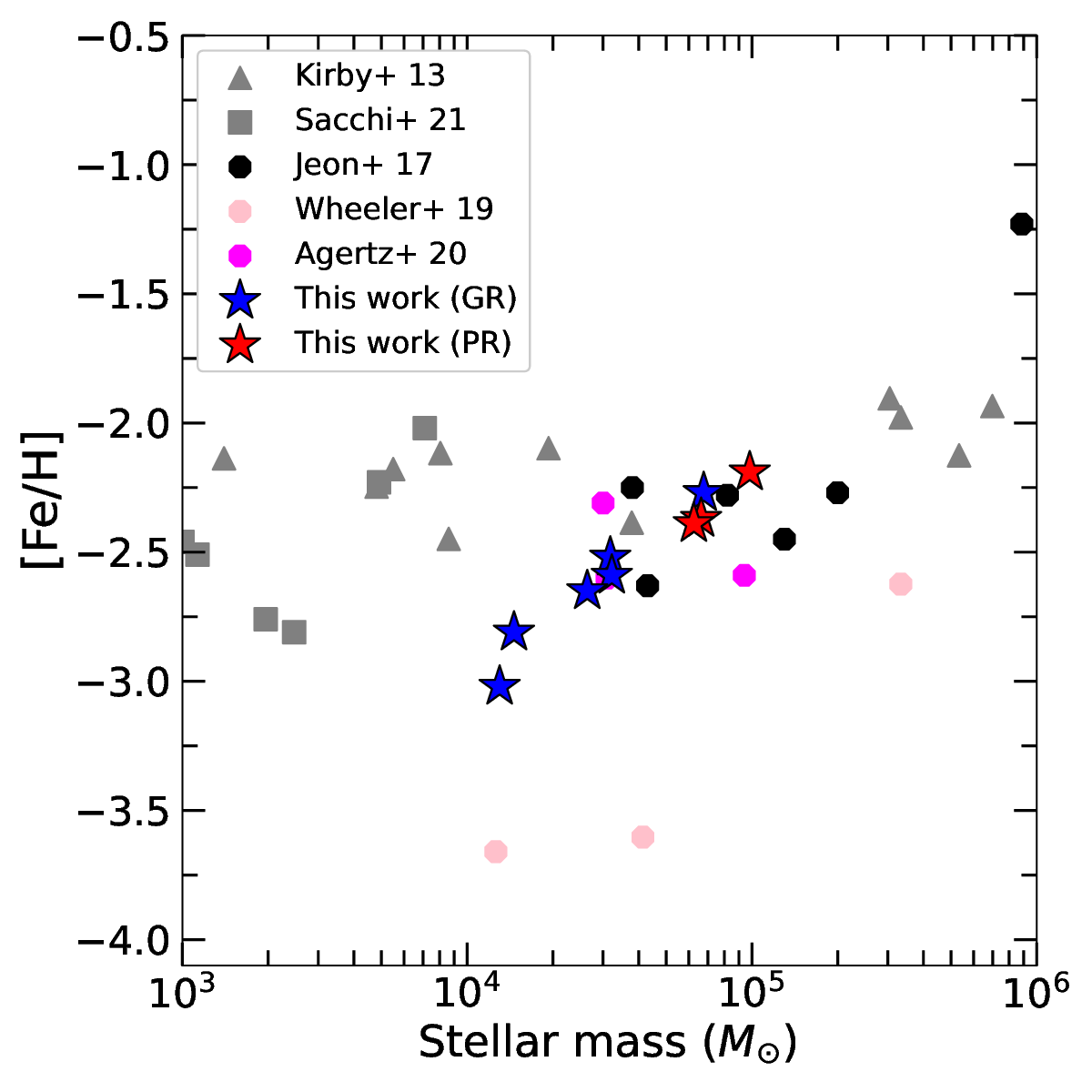}%
   \caption{The relationship between the stellar mass and the mean $\rm [Fe/H]$ of the simulated UFD analogs, in comparison with observations (\citealp{Kirby2013}; \citealp{Sachhi2021}) and other theoretical studies (\citealp{Jeon2017}; \citealp{Wheeler2019}; \citealp{Agertz2020}). The PR runs show that the metallicities of our simulated UFDs tend to agree well with those of observed MW satellites in the stellar mass range of $10^{4}\msun \lesssim M_{\rm *} \lesssim 10^{5}\msun$. However, reproducing the metallicity plateau in the lowest mass galaxy regime ($M_{\rm *} < 10^{4}\msun$), where $-3\lesssim\rm [Fe/H]\lesssim-2$ remains challenging for our simulations. While patchy reionization can contribute to increasing stellar metallicities, it also pushes them to the higher stellar mass regime ($M_{\rm *} \gtrsim 10^{4}\msun$).)   
   }
   \label{fig12}
\end{figure}

\begin{figure*}
  \centering
  \includegraphics[width=140mm]{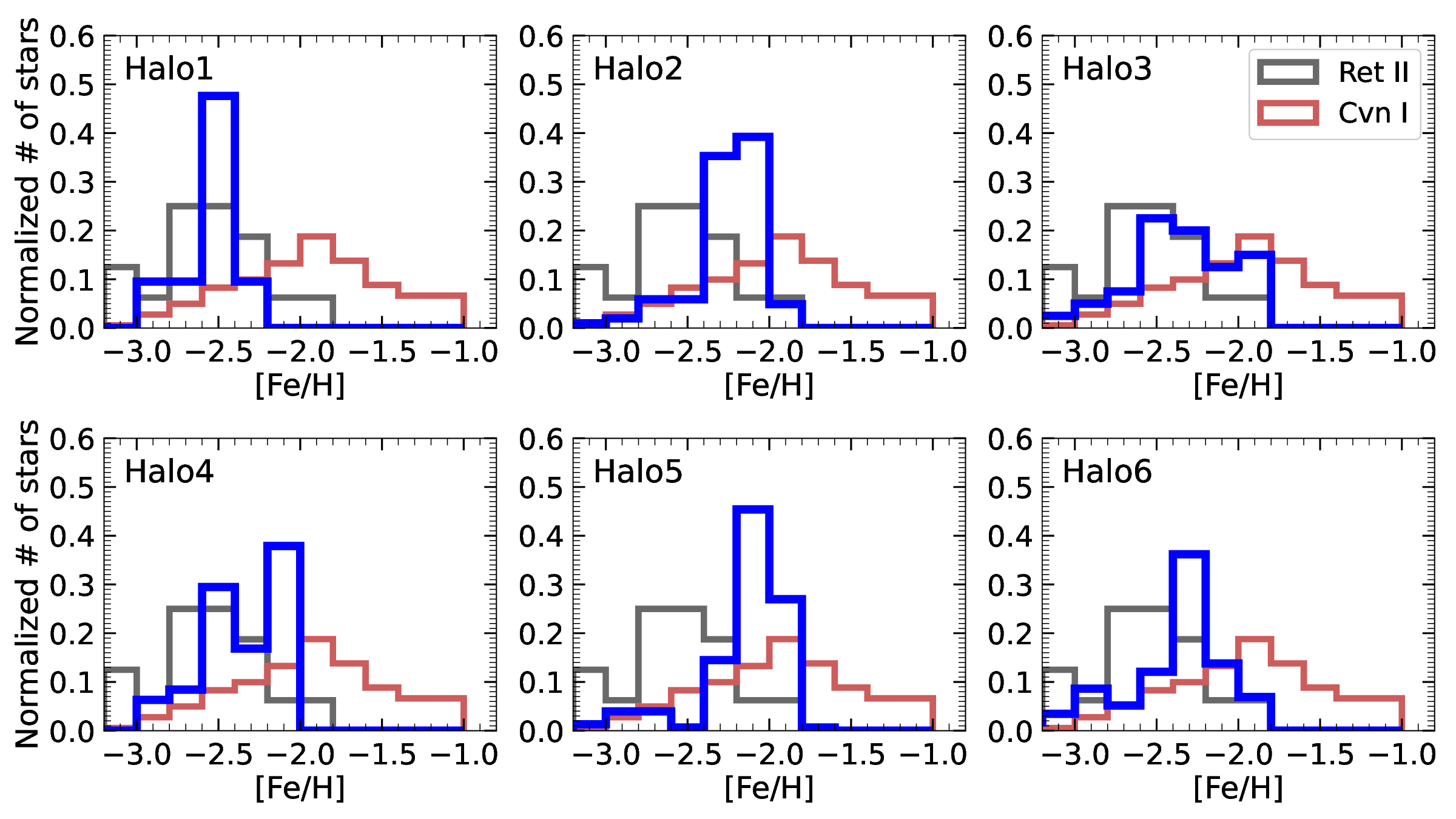}%
   \caption{The MDF of our simulated UFDs (shown in blue) in the PR runs adopting $z_{\rm t}=5.5$ is compared with that of Reticulum~II (gray), one of the Magellanic UFD satellites, and Canes Venatici~I (brown), in the form of a normalized histogram. Despite Reticulum II having a stellar mass lower by 1-2 orders of magnitude than the simulated galaxies (with $<M_{\ast}>\enspace=\enspace7.6\times10^{4}\msun$), its mean stellar metallicity ($\rm <[Fe/H]>\enspace= -2.46$) is similar to that of the simulated galaxies ($\rm <[Fe/H]>\enspace= -2.32$). On the other hand, Canes Venatici~I, which is about twice as massive as the simulated galaxy (blue), likely exhibits a wide range of metallicities, including stars with relatively high metallicity at $\rm [Fe/H]=-1$. Achieving such high metallicity stars is a challenging aspect of our study.}
   \label{fig13}
\end{figure*}

\subsubsection{Stellar metallicity} \label{sec:result_obs_metal}

Figure~\ref{fig12} presents a comparison of the estimated stellar mass and averaged [Fe/H] relation of our simulated UFDs, indicated both by blue and red filled stars, with those from other hydrodynamic simulations (\citealp{Jeon2017}; \citealp{Wheeler2019}; \citealp{Agertz2020}) and observations (\citealp{Kirby2013}; \citealp{Sachhi2021}). Our simulation results show that the estimated [Fe/H] values range from $\rm -3 \lesssim [Fe/H] \lesssim -2$ for stellar masses between $M_{\ast}=10^{4}\msun$ and $M_{\ast}=10^{5}\msun$. We find that the averaged [Fe/H] value of the runs with homogeneous reionization, $\rm <[Fe/H]>\enspace= -2.64$, is lower by 0.32 dex than that of the cases with patchy reionization. It is important to note that only three runs ({\sc Halo2-PR, Halo4-PR, Halo5-PR}) are considered for the patchy reionization cases since only these runs exhibit the impact of patchy reionization.

Theoretical studies tend to predict lower [Fe/H] values for UFDs with stellar masses in the range of $M_{\ast}\lesssim2\times10^4\msun$ compared to the observed values. In order to bridge the gap between observation and theory, several theoretical studies have been conducted. For example, \citet{Wheeler2019} suggested that this discrepancy in metallicity might be due to the neglect of Pop~III or a lack of environmental pre-enrichment. In addition, \citet{Jaacks2019} demonstrated that pre-enrichment by Pop~III is insufficient to raise the metallicity floor above $\rm [Fe/H] = -4$, particularly in low-density regions. On the other hand, \citet{Applebaum2021}, who adopted the same metallicity floor, successfully reproduced considerably more metal-enriched galaxies than \citet{Wheeler2019}. Furthermore, strong feedback from SNe may also contribute to the low metallicity, as \citet{Agertz2020} showed that SNe feedback could expel enriched gas out of the galaxy, thereby reducing the metallicity of the gas.

Our simulation results are also in line with other theoretical works, given that the predicted $\rm [Fe/H]$ values are lower than what is observed. For example, the estimated $\rm [Fe/H]$ for the smallest galaxy in our simulations, with $M_{\ast}\approx10^4\msun$, is around $\rm [Fe/H]=-2.8$, which is on average 0.5 dex lower than the observed values for UFDs with similar stellar masses. While implementing patchy reionization in our simulations can increase the stellar metallicities due to the formation of relatively high-metallicity stars during late star formation, it also leads to an increase in the stellar masses.

In Figure~\ref{fig13}, we compare the MDF of our simulated UFDs, shown as a blue histogram, with that of the observed UFDs, specifically Reticulum~II and Canes Venatici~I. These observed UFDs have estimated stellar masses of $M_{\ast}\sim10^{3}\msun$ and $M_{\ast}\sim2\times10^{5}\msun$, respectively. The metallicity measurements for Reticulum~II and Canes Venatici~I are provided by the Stellar Abundances for Galactic Archaeology (SAGA) database (\citealp{Suda2008}). Both the simulated UFDs and Reticulum~II consist mainly of metal-poor stars, spanning a metallicity range from $\rm [Fe/H]=-3.2$ to $\rm [Fe/H]=-1.7$. Despite the substantial difference in stellar mass between Reticulum~II and the simulated galaxies (with a mean stellar mass of $<M_{\ast}>\enspace=\enspace7.6\times10^{4}\msun$), the mean metallicity of member stars in Reticulum~II ($\rm <[Fe/H]>\enspace= -2.46$) is comparable to that of the simulated galaxies ($\rm <[Fe/H]>\enspace= -2.32$). As illustrated in Figure~\ref{fig11}, Reticulum~II, which is one of the Magellanic UFDs expected to show the effect of patchy reionization, is likely to have formed the majority of its stars ($\sim90\%$) before reionization, with only 10\% of its stars forming during the late star formation phase. On the contrary, Canes Venatici~I, which is approximately twice as massive as the simulated galaxy, is likely to display a wide range of metallicities, including stars with relatively high metallicity at $\rm [Fe/H]=-1.0$.

The formation of relatively high-metallicity stars ($\rm [Fe/H]\sim-1.0$) may not always occur at a later epoch when the global cosmic enrichment is achieved. Instead, during the rapid assembly process, even before reionization, high-metallicity stars can form from gas that has not had enough time to mix or diffuse into the surrounding gas. Note, however, that such a scenario is not demonstrated in our simulations. Furthermore, as demonstrated in Figure~\ref{fig7}, we find that stars formed during the late star formation phase tend to have metallicities similar to or lower than those formed before reionization. This is because even a weak patchy reionization can temporally suppress star formation between $z=7-6$, during which metals in the dense gas are dispersed by reionization, resulting in subsequent stars forming with lower metallicity.

\begin{figure}
  \centering
  \includegraphics[width=83mm]{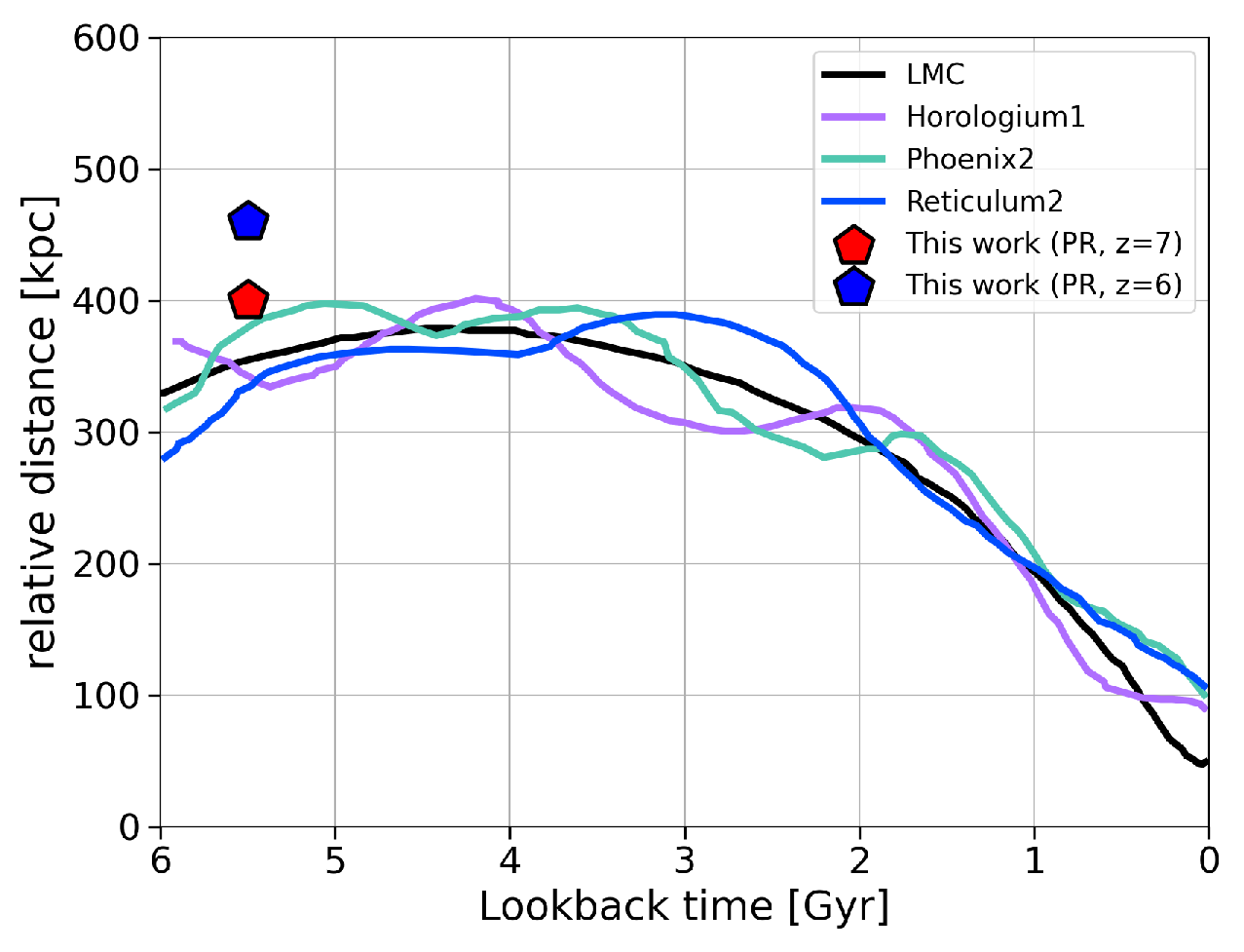}%
   \caption{The direct orbits of the Magellanic UFD satellites and the LMC from the center of the MW, reconstructed using Gaia's proper motion measurements (\citealp{Patel2020}). For comparison, the red pentagon represents the average distance of the simulated UFD analogs from the host halo at $z=7$, while the blue pentagon represents the average distance at $z=6$. Due to the lack of orbital histories of the Magellanic UFDs prior to 6 Gyr ago, it is difficult to determine the location of the observed UFDs during reionization. However, if the Magellanic UFDs were distant from the MW ($d >$ 400 kpc) at the end of reionization, they may have been exposed to a weaker local UV field from the MW progenitor, which could have allowed for star formation to continue after reionization.}
   \label{fig14}
\end{figure}

\subsubsection{Orbital histories of the observed UFDs}
\label{sec:result_obs_dist}
In order to investigate how far the simulated galaxy was from its host halo at the time of reionization, we show the average distance of the simulated galaxy with respect to its host halo at $z=7$ and $z=6$, marked as pentagon symbols in Figure~\ref{fig14}. We compare these distances with the orbital histories of Magellanic UFD satellites and the LMC itself over the past 6 Gyr. The solid lines in Figure~\ref{fig14} represent the reconstructed orbital histories of selected Magellanic satellites (Horologium~1, Phoenix~2, and Recticulum~2), derived from Gaia DR2 proper motion measurements, by taking into account the combined gravitational effects of the MW, LMC, and SMC (\citealp{Patel2020}). These results suggest that the LMC is currently on a first infall, long-period orbit and that its recent pericenter occurred 50 Myr ago, assuming an MW mass of $M_{\rm vir}\lesssim 1.5\times10^{12}\msun$. Moreover, the farthest distance from the MW over the last 6 Gyr is around 360 kpc. It should be noted that the unit of length used is physical rather than comoving units.

To make a more precise comparison with our simulations, it would be useful to have knowledge of the distances between the LMC and its associated satellites and the MW progenitor halo at earlier times than 6 Gyr. However, this is challenging as it requires a sophisticated orbital model that considers the mass evolution of MW-like galaxies, which have acquired roughly 80\% of their mass by 6 Gyr (e.g., \citealp{Santistevan2020}). Moreover, according to \citet{Patel2017}, results obtained from further orbit integration earlier than 6 Gyr may deviate from the predictions made by cosmological simulations. Despite this, we can anticipate that the LMC and its satellites were located far away during reionization since they appear to be currently undergoing their first infall (e.g., \citealp{Busha2011, Patel2017, Patel2020}). In particular, \citet{Patel2017} studied LMC analogs in Illustris simulations (e.g., \citealp{Vogelsberger2014, Nelson2015}) and calculated the cross time, which refers to the lookback time when the LMC initially crossed the physical $z=0$ virial radius of the MW while moving inward. Their analysis revealed that 40\% of LMC analogs had a cross time of less than 2 Gyr ago, while about 20\% had a crossing time of less than 4 Gyr ago. These results suggest that the LMC experienced its first infall relatively recently and was located far away from the MW progenitor halo at earlier times.

Based on the reconstructed orbital histories of Magellanic systems, there is a possibility that the UV intensity field used in our simulated galaxy is underestimated when compared to what the LMC satellites received. This is because the UV field in our simulation is mainly attributed to the MW-like host, and the contribution of the LMC is not adequately taken into account. If the LMC satellites were long-term satellites that were captured early on, their proximity to the LMC would have made the UV intensity from the LMC potentially significant. \citet{Patel2020} classified the satellites associated with the LMC into two categories based on the number of passages around the LMC, which depends on the masses of the MW and LMC and the inclusion of the SMC's contribution. Reticulum II and Phoenix II are believed to be recently captured systems as they completed one bound orbit around the MCs in the last 1 Gyr, while Horologium I is classified as a long-term satellite as it completed multiple passages around the MCs in the last 6 Gyr. Considering the above information, it is possible that the local UV field intensity by the LMC may not have had a significant effect on the recently captured LMC satellites because the LMC was relatively more distant at the time of reionization. However, in the case of long-term satellites like Horologium~I, the LMC might have contributed to quenching star formation due to its own potentially significant UV intensity.

\section{Summary and Conclusions} \label{sec:conclusion}
We have conducted a study to examine how patchy reionization affects the star formation histories (SFHs) and stellar metallicity of ultra-faint dwarfs (UFDs) by utilizing a set of cosmological hydrodynamic zoom-in simulations. Our study is motivated by recent findings by \citet{Sachhi2021}, who proposed that the contrasting SFHs between long-established Milky Way (MW) UFDs and recently entered ones could be attributed to the influence of patchy reionization. Specifically, the extended SFH observed in the Magellanic UFD satellites, which are estimated to have entered the MW's potential about $\sim$3.5 Gyr ago based on their orbital histories reconstructed using Gaia proper motion measurements, is thought to be a result of environmental factors, such as a weaker UV field during the reionization epoch compared to that experienced by the long-established MW satellites. Although patchy reionization could play a crucial role in shaping the formation and evolution of UFDs, the most common implementation of the reionization effect is simplistic, applying a uniform and instantaneous global heating to all galaxies within the simulation box without accounting for the relative distances between a target UFD and its surrounding galaxies.

We employ a novel method to account for the impact of local UV fields on a target UFD analog by surrounding galaxies, particularly its host galaxy. This method involves three steps to calculate the effects of patchy reionization. First, we conduct dark-matter-only simulations to identify a target UFD analog halo and track its distance from surrounding halos, as well as their halo masses. Second, we use the abundance-matching technique to determine the stellar masses of the surrounding halos. Third, we obtain synthetic galaxy spectra for the surrounding galaxies from Starburst99 to derive flux and subsequently calculate the photoionization and photoheating rates for hydrogen and helium atoms. We then perform cosmological hydrodynamic zoom-in simulations using these rates, which are provided in a table as a function of redshift. To clarify, our study involves a series of simulations that compare two models: one that employs homogeneous global reionization (GR) and another that uses a hybrid approach. The hybrid approach starts with patchy reionization and then switches to global reionization implementation below a transition redshift $z_{\rm t}$ (referred to as PR).

Our main findings are summarized as follows.

\begin{itemize}
\item {\sc Global reionization effect}
\begin{itemize}
    \item We confirm that reionization has a significant effect in suppressing star formation in the simulated UFDs. This is evidenced by the significant decrease in the maximum gas densities of all halos by two orders of magnitude from $z=7$ to $z=6$, which hinders the formation of new stars and results in no star formation in all halos below $z\approx7$.
    
    \item The amount of gas loss due to reionization depends on the halo mass at the onset of reionization. For instance, {\sc Halo5-GR}, the most massive halo at $z=7$, retains a gas mass of $M_{\rm gas}\sim$ 1.3 $\times$ 10$^{7}$$\msun$, which is three times larger than that of the other halos, while {\sc Halo2-GR} loses approximately 90\% of its gas mass ($M_{\rm gas}\sim2\times10^{6}\msun$ at $z=0$) between $z=6$ and $z=0$.
    
    \item Our results indicate that stars with higher metallicities ($\rm [Fe/H]\gtrsim-2$) are challenging to form due to the combined effects of supernova feedback and reionization. Moreover, we find that stars with extremely low metallicities ($\rm [Fe/H]\lesssim -5$) are formed through external metal enrichment.
\end{itemize}

\item {\sc Patchy reionization effect}
\begin{itemize}
    \item Most halos experience a temporary halt of star formation due to the patchy UV field during the PR runs, except for {\sc Halo5-PR}. However, the reduced intensity of patchy reionization, which is about two orders of magnitude lower than that in the GR scenario, allows the gas density to recovering to a level that enables star formation to resume in some PR cases.

    \item The occurrence of complete quenching or late star formation depends on the halo mass at the onset of reionization at $z=7$. For example, halos with higher masses, such as {\sc Halo2-PR}, {\sc Halo4-PR}, and {\sc Halo5-PR}, exhibit late star formation, while less massive halos, like {\sc Halo1-PR} and {\sc Halo3-PR}, experience complete quenching.

    \item The simulated halos, {\sc Halo2-PR}, {\sc Halo4-PR}, and {\sc Halo5-PR}, undergo late star formation and form a significant portion of their total stellar mass before $z=7$. Specifically, they form 60\%, 35\%, and 80\% of the total stellar mass, respectively. {\sc Halo5-PR} is unique in having a significantly longer star formation history, extending 550 Myr since $z=7$, compared to 280 Myr in {\sc Halo2-PR} and 180 Myr in {\sc Halo4-PR}.

    \item Multiple factors, such as the halo mass at the time of reionization, the duration of star formation, and the degree of star formation burstiness, influence the fraction of stars formed during late star formation.

    \item The in-situ stars formed through late star formation tend to have metallicities similar to or lower than the peak [Fe/H] of stars formed before reionization despite being formed at lower redshifts. This is due to the temporary quenching experienced by the simulated galaxies under the influence of a weak patchy reionization effect, which causes the dissipation of dense gas and the loss of associated metals.

    \item Our simulations indicate that by delaying the transition from PR to GR from $z_{\rm t}=5.8$ to $z_{\rm t}=5.5$, the SFHs are more extended with an average extension of 235 Myr. This prolonged period of star formation results in an increase in stellar mass, with a 63\% increase, observed in {\sc Halo4-PR}. Moreover, we find that the average metallicity of the UFD increases proportionally to the fraction of late star formation, causing the metallicity distribution function to shift toward higher metallicity by 0.2 dex.
\end{itemize}
\item {\sc Comparison with observations}
\begin{itemize}
    \item We find that $\tau_{\rm 90}$ is similar between Magellanic UFDs and {\sc PR-z5.5} runs, indicating that 90\% of stars are formed by $z\approx5$ in both cases. However, the shape of the SFHs differs between the two, with Magellanic UFDs forming stars earlier than the {\sc PR-z5.5} runs, with $\tau_{\rm 50}$ occurring 13.5 Gyr ago and 12.7 Gyr ago, respectively.
    
    \item Our simulations are consistent with the idea proposed by \citet{Sachhi2021} that patchy reionization could substantially extend the duration of star formation in UFDs. The quenching times of UFD analogs with extended SFHs in our PR runs adopting $z_{\rm t}=5.5$ are 460 Myr later than those in the GR runs, similar to the 600 Myr more recent quenching times of Magellanic UFDs compared to non-Magellanic UFDs. 
    
    \item Our simulations do not perfectly reproduce the absolute duration of late star formation observed in Magellanic UFDs. As such, the observed Magellanic UFDs exhibit a longer period of star formation, with a star formation cessation ($\rm SF_{\rm end}$) occurring 8.5 Gyr ago, whereas the PR runs indicate a mean $\rm SF_{\rm end}$ of 12.5 Gyr ago. This discrepancy in $\rm SF_{\rm end}$ could potentially be attributed to the comparatively milder reionization experienced by the Magellanic UFDs within their specific environment.
    
    \item Reproducing the high metallicity plateau of $\rm -2.7\lesssim[Fe/H]\lesssim-2.0$ below $M_{\ast}=2\times10^4\msun$, which is the mass range of Magellanic UFDs used for comparison, is challenging in our simulations. While patchy reionization can increase stellar metallicities by forming high-metallicity stars during late star formation, it also leads to an increase in stellar masses.
    
    \item The prolonged SFHs of the Magellanic UFDs with $\rm SF_{\rm end}=$ 8.5 Gyr ago could imply that the strength of the UV field they were subjected to might have been lower than the assumptions made in our PR runs. This discrepancy could suggest that the Magellanic UFDs were positioned farther away than the inferred distance of 400 kpc, based on reconstructed orbital histories at $z\approx0.7$.    
\end{itemize}
\end{itemize}


It is important to emphasize that the weakened UV background resulting from patchy reionization is not the sole explanation for the extended SFHs, and various factors can contribute to differing SFHs of dwarf galaxies. For instance, \citet{Rey2020} demonstrated that UFDs formed in halos with the same halo mass but varying assembly times can exhibit similar stellar masses while having distinct SFHs. To be specific, their study revealed that star formation can be reignited at later epochs ($z<1$) due to a significant increase in dynamical mass during these late periods. However, it is worth noting that the halos in their study have a mass of $M_{\rm vir}\approx3\times10^9\msun$ at $z=0$, which is 2-4 times more massive than those in our research. In a fair comparison, the least massive galaxy among their halos, with a mass of $M_{\rm vir}\approx1.4\times10^9\msun$ at $z=0$, similar to {\sc Halo5} in our study, also experiences complete quenching due to reionization.

As such, the mass of the halo is one of the crucial determinants of whether stars continue to form under the influence of reionization, leading to a minimum mass below which star formation is likely to be entirely quenched. This threshold varies from $M_{\rm vir}=3\times10^9\msun$ (e.g., \citealp{Jeon2017, Rey2020}) to $M_{\rm vir}=7\times10^9\msun$ (e.g., \citealp{Fitts2017, Wheeler2019}. However, explaining the extended SFHs of the observed Magellanic UFDs by adopting more massive halos presents a significant challenge. This is primarily because the estimated stellar mass of the Magellanic UFDs, which is on the order of a few $M_{\ast}\sim10^3\msun$, is too small to be hosted in massive halos with $M_{\rm vir}\gtrsim3\times10^9\msun$. Theoretical predictions generally suggest that $M_{\rm vir}\lesssim10^9\msun$ is required to produce a stellar mass of $M_{\ast}\sim10^3\msun$. However, as mentioned earlier, within this mass range, the effects of reionization are likely to result in complete quenching when adopting the conventional UV background ({\sc HM2012}). Therefore, the extended SFHs observed in the Magellanic UFDs are more likely a consequence of encountering a weak UV field during the reionization epoch rather than being hosted by highly massive halos. Moreover, if massive halos are indeed the cause, an even weaker UV field than what we have adopted would be required to generate the more prolonged SFHs observed in the Magellanic UFDs.

One possible approach to diminish the intensity of the UV field is to consider an alternative escape fraction of ionizing photons from galaxies, rather than the currently adopted value of $f_{\rm esc}=0.3$. A lower escape fraction would result in more extended SFHs (refer to Appendix~\ref{appendix_c}). It has been suggested that there may be a correlation between star formation rate and escape fraction and that the escape fraction may evolve with redshift (e.g., \citealp{Kunlen2012}). According to \citet{Ma2020}, for instance, the average escape fraction of ionizing photons from galaxies increases with halo mass for $M_{\rm vir}=10^8-10^{9.5}\msun$, remains roughly constant, $f_{\rm esc}=0.1-0.2$, for $M_{\rm vir}=10^{9.5}-10^{11}\msun$, and then drops for $M_{\rm vir}=10^{11}\msun$ due to dust attenuation, with an $f_{\rm esc}$ of less than 0.1 at the massive end.

Finally, we would like to emphasize that we are not claiming that our patchy reionization model surpasses the HM2012 model, primarily because we did not incorporate the influence of galaxies on a large scale. However, it is important to note that previous simulations, especially those concentrated on UFD galaxy formation, have typically employed a flash-like reionization approach that affects all gas particles, resulting in an abrupt cessation of star formation within UFDs. In contrast, our approach in this study considers the unique environments of the simulated UFD analogs, allowing us to test the patchy reionization model as a potential explanation for the extended SFHs observed in Magellanic UFDs. To further validate our predictions regarding the effect of patchy reionization on the SFHs of observed UFDs, additional SFHs for satellite galaxies would be beneficial. The forthcoming SFHs for Magellanic satellites, obtained from deep HST and JWST imaging, will help identify differences between the SFHs of Magellanic UFDs
and long-standing satellites of the MW. Additionally, spectroscopic surveys could provide more accurate constraints on metallicities and enable us to estimate more precise ages of stars.

\section*{Acknowledgements}
We would like to thank Gurtina Besla for providing valuable comments. We are grateful to Volker Springel, Joop Schaye, and Claudio Dalla
Vecchia for permission to use their versions of \textsc{gadget}.
J.~L. and M.~J. are supported by the National Research Foundation (NRF) grant No. 2021R1A2C109491713, funded by the Korean government (MSIT). Y.C. acknowledges support from NASA grant No. HST-GO-16293.

\clearpage

\appendix
\section{Results of the UFD analogs with two different reionization implementations}\label{appendix_a}

\setcounter{table}{0}
\renewcommand{\thetable}{A\arabic{table}}
\renewcommand*{\theHtable}{\thetable}

\begin{deluxetable*}{cccccccccc}[b]
\tablecaption{Physical quantities of the simulated UFD analogs at $z=3$ (see, Section~\ref{sec:result_zt})\label{table3}}
\tabletypesize{\small}
\tablewidth{0pt}
\tablehead{
\colhead{Halo} & \colhead{$M_{\rm vir}$} & \colhead{$M_{\ast}$} & \colhead{$\rm <[Fe/H]>$} & \colhead{$M_{\rm gas}$} & \colhead{$z_{\rm t}$} & \colhead{$z_{\rm quen}$} & \colhead{$z_{\rm sf,start}$} & \colhead{$z_{\rm sf,end}$} & \colhead{$t_{\rm dur}$} \\
\colhead{} & \colhead{$[10^8\msun]$} & \colhead{$[10^4\msun]$} & \colhead{-} & \colhead{$[10^6\msun]$} & \colhead{-} & \colhead{-} & \colhead{-} & \colhead{-} & \colhead{[Myr]} 
}

\startdata
{\sc Halo1-GR}  & 1.97  & 1.30 & -3.02  & 6.52  & - & 7.036 & - & - & - \\
{\sc Halo1-PR-z5.8}  & 1.97  & 1.36 & -2.98  & 6.89  & 5.8 & 7.036 & 6.639 & 6.639 & - \\
{\sc Halo1-PR-z5.7}  & 1.97  & 1.36 & -2.98  & 6.93  & 5.7 & 7.036 & 6.639 & 6.639 & - \\
{\sc Halo1-PR-z5.6}  & 1.97  & 1.36 & -2.98  & 6.96  & 5.6 & 7.036 & 6.639 & 6.639 & - \\
{\sc Halo1-PR-z5.5}  & 1.97  & 1.36 & -2.98  & 7.02  & 5.5 & 7.036 & 6.639 & 6.639 & - \\
\hline
{\sc Halo2-GR}  & 3.17   & 3.18 & -2.52 & 2.03  & - & 6.998 & - & - & - \\
{\sc Halo2-PR-z5.8}  & 3.17   & 5.07 & -2.45 & 2.16  & 5.8 & 6.953 & 6.034 & 4.933 & 277 \\
{\sc Halo2-PR-z5.7}  & 3.17   & 5.98 & -2.40 & 2.20  & 5.7 & 6.953 & 6.034 & 4.876 & 295 \\
{\sc Halo2-PR-z5.6}  & 3.17   & 5.85 & -2.40 & 2.22  & 5.6 & 6.953 & 6.034 & 4.885 & 292 \\
{\sc Halo2-PR-z5.5}  & 3.17   & 6.63 & -2.37 & 2.22  & 5.5 & 6.953 & 6.034 & 4.909 & 284 \\
\hline
{\sc Halo3-GR}  & 3.89   & 2.64 & -2.65 & 2.96 & - & 7.106 & - & - & - \\
{\sc Halo3-PR-z5.8}  & 3.89  & 2.64 & -2.65  & 3.32  & 5.8 & 7.106 & - & - & - \\
{\sc Halo3-PR-z5.7}  & 3.89  & 2.64 & -2.65  & 3.39  & 5.7 & 7.106 & - & - & - \\
{\sc Halo3-PR-z5.6}  & 3.89  & 2.64 & -2.65  & 3.44  & 5.6 & 7.106 & - & - & - \\
{\sc Halo3-PR-z5.5}  & 3.89  & 2.64 & -2.65  & 3.46  & 5.5 & 7.106 & - & - & - \\
\hline
{\sc Halo4-GR}  & 3.87   & 1.46 & -2.81 & 4.07  & - & 7.037 & - & - & - \\
{\sc Halo4-PR-z5.8}  & 3.87   & 3.84 & -2.59 & 4.26  & 5.8 & 7.037 & 6.223 & 5.430 & 174 \\
{\sc Halo4-PR-z5.7}  & 3.87   & 4.82 & -2.48 & 4.31  & 5.7 & 7.037 & 6.223 & 4.831 & 346 \\
{\sc Halo4-PR-z5.6}  & 3.87   & 4.94 & -2.48 & 4.27  & 5.6 & 7.037 & 6.223 & 5.313 & 205 \\
{\sc Halo4-PR-z5.5}  & 3.87   & 6.25 & -2.39 & 4.35  & 5.5 & 7.037 & 6.223 & 4.639 & 411 \\
\hline
{\sc Halo5-GR}  & 5.35   & 6.77 & -2.27 & 7.68 & - & 6.917 & - & - & - \\
{\sc Halo5-PR-z5.8}  & 5.35  & 8.38 & -2.22  & 9.66  & 5.8 & 6.859 & 6.630 & 4.454 & 550 \\
{\sc Halo5-PR-z5.7}  & 5.35  & 8.77 & -2.21  & 9.19  & 5.7 & 6.859 & 6.630 & 5.113 & 332 \\
{\sc Halo5-PR-z5.6}  & 5.35  & 9.03 & -2.20  & 10.04  & 5.6 & 6.859 & 6.630 & 4.638 & 483 \\
{\sc Halo5-PR-z5.5}  & 5.35  & 9.82 & -2.19  & 12.91  & 5.5 & 6.859 & 6.630 & 3.566 & 974 \\
\hline
{\sc Halo6-GR}  & 4.95   & 3.22 & -2.59 & 5.75  & - & 6.895 & - & - & - \\
{\sc Halo6-PR-z5.8}  & 4.95   & 3.74 & -2.69 & 6.47  & 5.8 & 7.040 & - & - & - \\
{\sc Halo6-PR-z5.7}  & 4.95   & 3.74 & -2.69 & 6.90  & 5.7 & 7.040 & - & - & - \\
{\sc Halo6-PR-z5.6}  & 4.95   & 3.80 & -2.69 & 6.64  & 5.6 & 6.990 & - & - & - \\
{\sc Halo6-PR-z5.5}  & 4.95   & 3.80 & -2.69 & 6.78  & 5.5 & 6.990 & - & - & - \\
\enddata
\tablecomments{Column (1): the name of the halo. Column (2): viral mass (in units of $10^8\msun$). Column (3): stellar mass  (in $10^4\msun$). Column (4): average stellar iron-to-hydrogen ratios. Column (5): gas mass (in units of $10^6\msun$). Column (6): transition time. Column (7): the time that the simulated UFDs are quenched by reionization at $z\sim7$. Column (8): the time of beginning late star formation. Column (9): the time of completing late star formation. Column (10): duration of late star formation.}
\end{deluxetable*}
\clearpage

\onecolumngrid

\renewcommand\thefigure{\thesection \arabic{figure}}    
\section{The strength of patchy reionization on a large scale}\label{appendix_b}
\setcounter{figure}{0}
\begin{figure*}[hbp]
\centering
  \includegraphics[width=160mm]{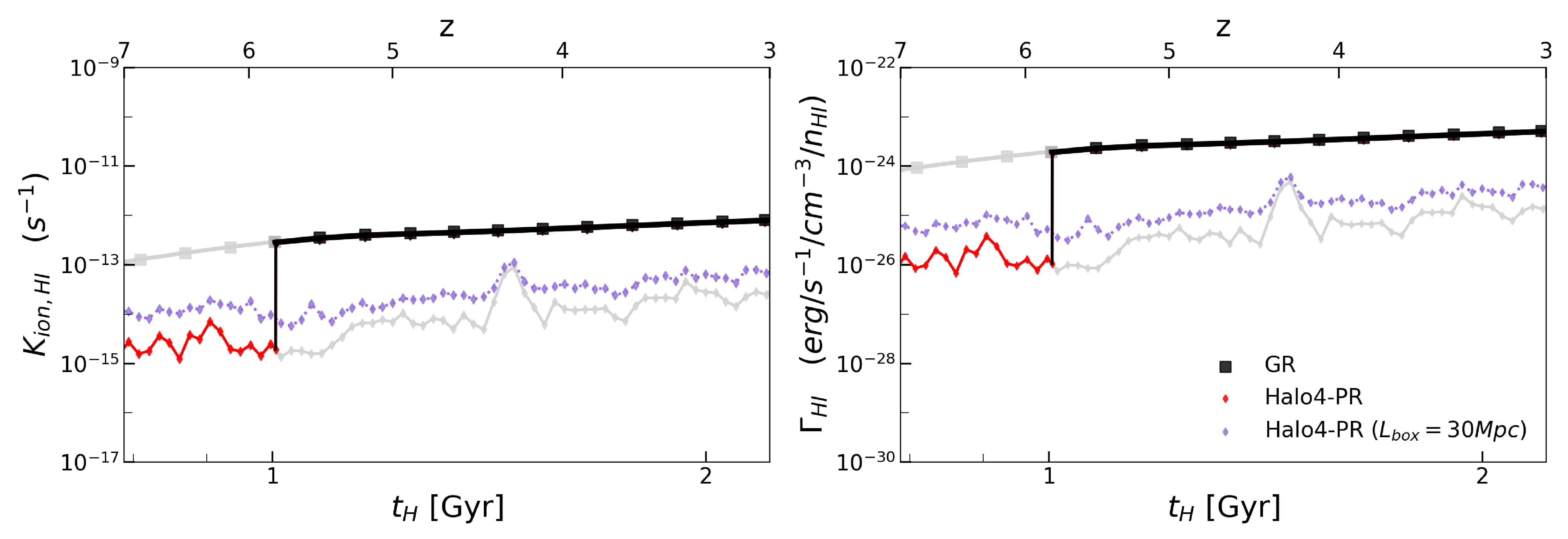}%
   \caption{The figure is similar to Figure~\ref{fig5}, but it specifically illustrates the hydrogen photoionization rate (left) and photoheating rate (right) produced by galaxies within a large box, spanning a linear size of 30 comoving Mpc. In this region, three massive galaxies (with $M_{\rm vir}\gtrsim10^{13}\msun$ at $z=0$) predominantly ($\gtrsim80\%$) influence the simulated UFD analog, {\sc Halo~4}. Despite the substantial mass of these halos and their significant contribution of UV photons to {\sc Halo~4}, it is noteworthy that during the reionization epoch, these three halos were, on average, five times farther from {\sc Halo~4} compared to the distance between {\sc Halo~4} and the MW-like halo. Consequently, the total level of reionization by all galaxies within the simulation box is elevated by a factor of 5-10 in comparison to what the Milky Way-mass halo imposes on the target UFDs. However, it remains lower by a factor of 20-50 compared to the intensity of GR.}
\end{figure*}

\renewcommand\thefigure{\thesection \arabic{figure}}    
\section{The effect of escape fraction on SFHs}\label{appendix_c}
\setcounter{figure}{0}
\begin{figure*}[hbp]
\centering
  \includegraphics[width=80mm]{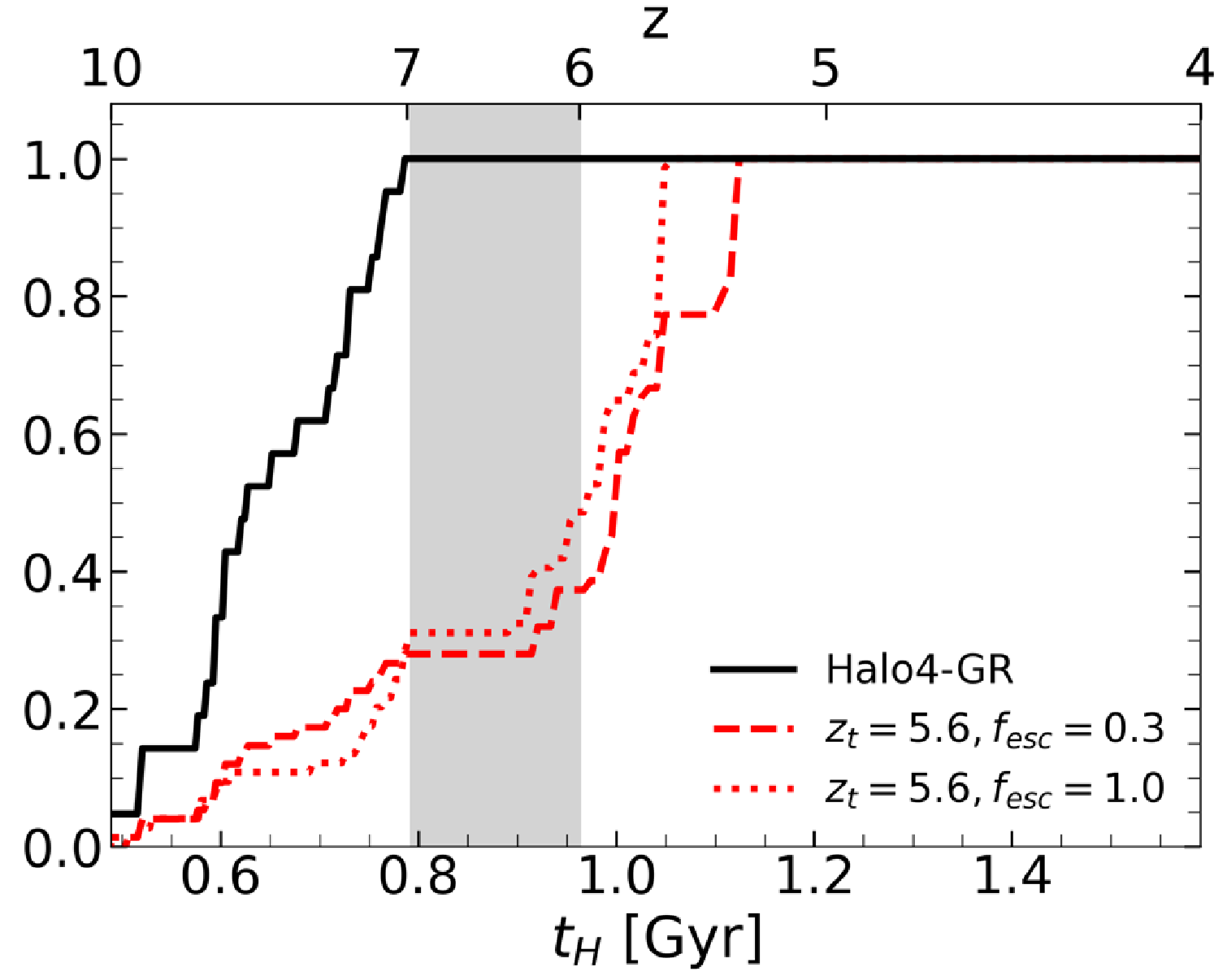}%
   \caption{The SFHs of {\sc Halo~4} by employing different escape fractions. When the escape fraction decreases from 100\% (dotted line) to 30\% (dashed line), the duration of the SFH extends by approximately 70 Myrs, indicating that more extended SFHs can be attained with a lower escape fraction. We find that the stellar masses are similar between the two cases, but the run with $f_{\rm esc}=1.0$ has 40\% less gas mass compared to the run adopting $f_{\rm esc}=0.3$.}
\end{figure*}
\bibliography{myrefs}{}

\begin{thebibliography}{}
\expandafter\ifx\csname natexlab\endcsname\relax\def\natexlab#1{#1}\fi
\providecommand{\url}[1]{\href{#1}{#1}}
\providecommand{\dodoi}[1]{doi:~\href{http://doi.org/#1}{\nolinkurl{#1}}}
\providecommand{\doeprint}[1]{\href{http://ascl.net/#1}{\nolinkurl{http://ascl.net/#1}}}
\providecommand{\doarXiv}[1]{\href{https://arxiv.org/abs/#1}{\nolinkurl{https://arxiv.org/abs/#1}}}

\bibitem[{{Agertz} \& {Kravtsov}(2015)}]{Agertz2015}
{Agertz}, O., \& {Kravtsov}, A.~V. 2015, ApJ, 804, 18,
  \dodoi{10.1088/0004-637X/804/1/18}

\bibitem[{{Agertz} {et~al.}(2020){Agertz}, {Pontzen}, {Read}, {Rey}, {Orkney},
  {Rosdahl}, {Teyssier}, {Verbeke}, {Kretschmer}, \& {Nickerson}}]{Agertz2020}
{Agertz}, O., {Pontzen}, A., {Read}, J.~I., {et~al.} 2020, MNRAS, 491, 1656,
  \dodoi{10.1093/mnras/stz3053}

\bibitem[{{Applebaum} {et~al.}(2021){Applebaum}, {Brooks}, {Christensen},
  {Munshi}, {Quinn}, {Shen}, \& {Tremmel}}]{Applebaum2021}
{Applebaum}, E., {Brooks}, A.~M., {Christensen}, C.~R., {et~al.} 2021, \apj,
  906, 96, \dodoi{10.3847/1538-4357/abcafa}

\bibitem[{{Barris} \& {Tonry}(2006)}]{Barris2006}
{Barris}, B.~J., \& {Tonry}, J.~L. 2006, ApJ, 637, 427, \dodoi{10.1086/498292}

\bibitem[{{Becker} {et~al.}(2015){Becker}, {Bolton}, {Madau}, {Pettini},
  {Ryan-Weber}, \& {Venemans}}]{Becker2015}
{Becker}, G.~D., {Bolton}, J.~S., {Madau}, P., {et~al.} 2015, \mnras, 447,
  3402, \dodoi{10.1093/mnras/stu2646}

\bibitem[{{Becker} {et~al.}(2001){Becker}, {Fan}, {White}, {Strauss},
  {Narayanan}, {Lupton}, {Gunn}, {Annis}, {Bahcall}, {Brinkmann}, {Connolly},
  {Csabai}, {Czarapata}, {Doi}, {Heckman}, {Hennessy}, {Ivezi{\'c}}, {Knapp},
  {Lamb}, {McKay}, {Munn}, {Nash}, {Nichol}, {Pier}, {Richards}, {Schneider},
  {Stoughton}, {Szalay}, {Thakar}, \& {York}}]{Becker2001}
{Becker}, R.~H., {Fan}, X., {White}, R.~L., {et~al.} 2001, \aj, 122, 2850,
  \dodoi{10.1086/324231}

\bibitem[{{Behroozi} {et~al.}(2013){Behroozi}, {Wechsler}, \&
  {Conroy}}]{Behroozi2013}
{Behroozi}, P.~S., {Wechsler}, R.~H., \& {Conroy}, C. 2013, ApJ, 770, 57,
  \dodoi{10.1088/0004-637X/770/1/57}

\bibitem[{{Bose} {et~al.}(2018){Bose}, {Deason}, \& {Frenk}}]{Bose18}
{Bose}, S., {Deason}, A.~J., \& {Frenk}, C.~S. 2018, \apj, 863, 123,
  \dodoi{10.3847/1538-4357/aacbc4}

\bibitem[{{Bouwens} {et~al.}(2015){Bouwens}, {Illingworth}, {Oesch}, {Trenti},
  {Labb{\'e}}, {Bradley}, {Carollo}, {van Dokkum}, {Gonzalez}, {Holwerda},
  {Franx}, {Spitler}, {Smit}, \& {Magee}}]{Bouwens15b}
{Bouwens}, R.~J., {Illingworth}, G.~D., {Oesch}, P.~A., {et~al.} 2015, \apj,
  803, 34, \dodoi{10.1088/0004-637X/803/1/34}

\bibitem[{{Brown} {et~al.}(2014){Brown}, {Tumlinson}, {Geha}, {Simon},
  {Vargas}, {VandenBerg}, {Kirby}, {Kalirai}, {Avila}, {Gennaro}, {Ferguson},
  {Mu{\~n}oz}, {Guhathakurta}, \& {Renzini}}]{Brown2014}
{Brown}, T.~M., {Tumlinson}, J., {Geha}, M., {et~al.} 2014, \apj, 796, 91,
  \dodoi{10.1088/0004-637X/796/2/91}

\bibitem[{{Busha} {et~al.}(2011){Busha}, {Marshall}, {Wechsler}, {Klypin}, \&
  {Primack}}]{Busha2011}
{Busha}, M.~T., {Marshall}, P.~J., {Wechsler}, R.~H., {Klypin}, A., \&
  {Primack}, J. 2011, \apj, 743, 40, \dodoi{10.1088/0004-637X/743/1/40}

\bibitem[{{Choi} {et~al.}(2020){Choi}, {Dalcanton}, {Williams}, {Skillman},
  {Fouesneau}, {Gordon}, {Sandstrom}, {Weisz}, \& {Gilbert}}]{Choi2020}
{Choi}, Y., {Dalcanton}, J.~J., {Williams}, B.~F., {et~al.} 2020, \apj, 902,
  54, \dodoi{10.3847/1538-4357/abb467}

\bibitem[{{Choudhury} {et~al.}(2015){Choudhury}, {Puchwein}, {Haehnelt}, \&
  {Bolton}}]{Choudhury2015}
{Choudhury}, T.~R., {Puchwein}, E., {Haehnelt}, M.~G., \& {Bolton}, J.~S. 2015,
  \mnras, 452, 261, \dodoi{10.1093/mnras/stv1250}

\bibitem[{{Dalla Vecchia} \& {Schaye}(2012)}]{Vecchia2012}
{Dalla Vecchia}, C., \& {Schaye}, J. 2012, MNRAS, 426, 140,
  \dodoi{10.1111/j.1365-2966.2012.21704.x}

\bibitem[{{Dixon} {et~al.}(2018){Dixon}, {Iliev}, {Gottl{\"o}ber}, {Yepes},
  {Knebe}, {Libeskind}, \& {Hoffman}}]{Dixon18}
{Dixon}, K.~L., {Iliev}, I.~T., {Gottl{\"o}ber}, S., {et~al.} 2018, \mnras,
  477, 867, \dodoi{10.1093/mnras/sty494}

\bibitem[{{Fan et al.}(2006)}]{Fan2006}
{Fan et al.} 2006, AJ, 131, 1203, \dodoi{10.1086/500296}

\bibitem[{{Faucher-Gigu{\`e}re} {et~al.}(2009){Faucher-Gigu{\`e}re}, {Lidz},
  {Zaldarriaga}, \& {Hernquist}}]{Faucher2009}
{Faucher-Gigu{\`e}re}, C.-A., {Lidz}, A., {Zaldarriaga}, M., \& {Hernquist}, L.
  2009, \apj, 703, 1416, \dodoi{10.1088/0004-637X/703/2/1416}

\bibitem[{{Ferland} {et~al.}(1998){Ferland}, {Korista}, {Verner}, {Ferguson},
  {Kingdon}, \& {Verner}}]{Ferland1998}
{Ferland}, G.~J., {Korista}, K.~T., {Verner}, D.~A., {et~al.} 1998, PASP, 110,
  761, \dodoi{10.1086/316190}

\bibitem[{{Fitts} {et~al.}(2017){Fitts}, {Boylan-Kolchin}, {Elbert}, {Bullock},
  {Hopkins}, {O{\~n}orbe}, {Wetzel}, {Wheeler}, {Faucher-Gigu{\`e}re},
  {Kere{\v{s}}}, {Skillman}, \& {Weisz}}]{Fitts2017}
{Fitts}, A., {Boylan-Kolchin}, M., {Elbert}, O.~D., {et~al.} 2017, \mnras, 471,
  3547, \dodoi{10.1093/mnras/stx1757}

\bibitem[{{F{\"o}rster} {et~al.}(2006){F{\"o}rster}, {Wolf}, {Podsiadlowski},
  \& {Han}}]{Forster2006}
{F{\"o}rster}, F., {Wolf}, C., {Podsiadlowski}, P., \& {Han}, Z. 2006, MNRAS,
  368, 1893, \dodoi{10.1111/j.1365-2966.2006.10258.x}

\bibitem[{{Garrison-Kimmel} {et~al.}(2019){Garrison-Kimmel}, {Wetzel},
  {Hopkins}, {Sanderson}, {El-Badry}, {Graus}, {Chan}, {Feldmann},
  {Boylan-Kolchin}, {Hayward}, {Bullock}, {Fitts}, {Samuel}, {Wheeler},
  {Kere{\v{s}}}, \& {Faucher-Gigu{\`e}re}}]{Garrison19}
{Garrison-Kimmel}, S., {Wetzel}, A., {Hopkins}, P.~F., {et~al.} 2019, \mnras,
  489, 4574, \dodoi{10.1093/mnras/stz2507}

\bibitem[{{Greif} {et~al.}(2009){Greif}, {Johnson}, {Klessen}, \&
  {Bromm}}]{Greif2009}
{Greif}, T.~H., {Johnson}, J.~L., {Klessen}, R.~S., \& {Bromm}, V. 2009, MNRAS,
  399, 639, \dodoi{10.1111/j.1365-2966.2009.15336.x}

\bibitem[{{Gutcke} {et~al.}(2022){Gutcke}, {Pakmor}, {Naab}, \&
  {Springel}}]{Gutcke2022}
{Gutcke}, T.~A., {Pakmor}, R., {Naab}, T., \& {Springel}, V. 2022, \mnras, 513,
  1372, \dodoi{10.1093/mnras/stac867}

\bibitem[{{Haardt} \& {Madau}(2012)}]{Madau2012}
{Haardt}, F., \& {Madau}, P. 2012, \apj, 746, 125,
  \dodoi{10.1088/0004-637X/746/2/125}

\bibitem[{{Hahn} \& {Abel}(2011)}]{Hahn2011}
{Hahn}, O., \& {Abel}, T. 2011, MNRAS, 415, 2101,
  \dodoi{10.1111/j.1365-2966.2011.18820.x}

\bibitem[{{Heger} \& {Woosley}(2002)}]{Heger2002}
{Heger}, A., \& {Woosley}, S.~E. 2002, ApJ, 567, 532, \dodoi{10.1086/338487}

\bibitem[{{Heger} \& {Woosley}(2010)}]{Heger2010}
---. 2010, ApJ, 724, 341, \dodoi{10.1088/0004-637X/724/1/341}

\bibitem[{{Hopkins} {et~al.}(2020){Hopkins}, {Grudi{\'c}}, {Wetzel},
  {Kere{\v{s}}}, {Faucher-Gigu{\`e}re}, {Ma}, {Murray}, \&
  {Butcher}}]{Hopkins2020}
{Hopkins}, P.~F., {Grudi{\'c}}, M.~Y., {Wetzel}, A., {et~al.} 2020, MNRAS, 491,
  3702, \dodoi{10.1093/mnras/stz3129}

\bibitem[{{Jaacks} {et~al.}(2019){Jaacks}, {Finkelstein}, \&
  {Bromm}}]{Jaacks2019}
{Jaacks}, J., {Finkelstein}, S.~L., \& {Bromm}, V. 2019, \mnras, 488, 2202,
  \dodoi{10.1093/mnras/stz1529}

\bibitem[{{Jeon} {et~al.}(2017){Jeon}, {Besla}, \& {Bromm}}]{Jeon2017}
{Jeon}, M., {Besla}, G., \& {Bromm}, V. 2017, \apj, 848, 85,
  \dodoi{10.3847/1538-4357/aa8c80}

\bibitem[{{Khaire} {et~al.}(2016){Khaire}, {Srianand}, {Choudhury}, \&
  {Gaikwad}}]{Khaire16}
{Khaire}, V., {Srianand}, R., {Choudhury}, T.~R., \& {Gaikwad}, P. 2016,
  \mnras, 457, 4051, \dodoi{10.1093/mnras/stw192}

\bibitem[{{Kimm} \& {Cen}(2014)}]{Kimm2014}
{Kimm}, T., \& {Cen}, R. 2014, \apj, 788, 121,
  \dodoi{10.1088/0004-637X/788/2/121}

\bibitem[{{Kirby} {et~al.}(2013){Kirby}, {Cohen}, {Guhathakurta}, {Cheng},
  {Bullock}, \& {Gallazzi}}]{Kirby2013}
{Kirby}, E.~N., {Cohen}, J.~G., {Guhathakurta}, P., {et~al.} 2013, ApJ, 779,
  102, \dodoi{10.1088/0004-637X/779/2/102}

\bibitem[{{Komatsu} {et~al.}(2011){Komatsu}, {Smith}, {Dunkley}, {Bennett},
  {Gold}, {Hinshaw}, {Jarosik}, {Larson}, {Nolta}, {Page}, \&
  {Spergel}}]{Komatsu2011}
{Komatsu}, E., {Smith}, K.~M., {Dunkley}, J., {et~al.} 2011, ApJS, 192, 18,
  \dodoi{10.1088/0067-0049/192/2/18}

\bibitem[{{Kuhlen} \& {Faucher-Gigu{\`e}re}(2012)}]{Kunlen2012}
{Kuhlen}, M., \& {Faucher-Gigu{\`e}re}, C.-A. 2012, \mnras, 423, 862,
  \dodoi{10.1111/j.1365-2966.2012.20924.x}

\bibitem[{{Leitherer} {et~al.}(1999){Leitherer}, {Schaerer}, {Goldader},
  {Delgado}, {Robert}, {Kune}, {de Mello}, {Devost}, \&
  {Heckman}}]{Leitherer1999}
{Leitherer}, C., {Schaerer}, D., {Goldader}, J.~D., {et~al.} 1999, ApJS, 123,
  3, \dodoi{10.1086/313233}

\bibitem[{{Ma} {et~al.}(2016){Ma}, {Hopkins}, {Kasen}, {Quataert},
  {Faucher-Gigu{\`e}re}, {Kere{\v{s}}}, {Murray}, \& {Strom}}]{Ma2016}
{Ma}, X., {Hopkins}, P.~F., {Kasen}, D., {et~al.} 2016, \mnras, 459, 3614,
  \dodoi{10.1093/mnras/stw941}

\bibitem[{{Ma} {et~al.}(2020){Ma}, {Quataert}, {Wetzel}, {Hopkins},
  {Faucher-Gigu{\`e}re}, \& {Kere{\v{s}}}}]{Ma2020}
{Ma}, X., {Quataert}, E., {Wetzel}, A., {et~al.} 2020, \mnras, 498, 2001,
  \dodoi{10.1093/mnras/staa2404}

\bibitem[{{Marigo}(2001)}]{Marigo2001}
{Marigo}, P. 2001, \aap, 370, 194, \dodoi{10.1051/0004-6361:20000247}

\bibitem[{{Martin} {et~al.}(2007){Martin}, {Ibata}, {Chapman}, {Irwin}, \&
  {Lewis}}]{Martin2007}
{Martin}, N.~F., {Ibata}, R.~A., {Chapman}, S.~C., {Irwin}, M., \& {Lewis},
  G.~F. 2007, \mnras, 380, 281, \dodoi{10.1111/j.1365-2966.2007.12055.x}

\bibitem[{{McConnachie}(2012)}]{McConnachie2012}
{McConnachie}, A.~W. 2012, AJ, 144, 4, \dodoi{10.1088/0004-6256/144/1/4}

\bibitem[{{McGreer} {et~al.}(2015){McGreer}, {Mesinger}, \&
  {D'Odorico}}]{McGreer2015}
{McGreer}, I.~D., {Mesinger}, A., \& {D'Odorico}, V. 2015, \mnras, 447, 499,
  \dodoi{10.1093/mnras/stu2449}

\bibitem[{{Mesinger} {et~al.}(2015){Mesinger}, {Aykutalp}, {Vanzella},
  {Pentericci}, {Ferrara}, \& {Dijkstra}}]{Mesinger2015}
{Mesinger}, A., {Aykutalp}, A., {Vanzella}, E., {et~al.} 2015, \mnras, 446,
  566, \dodoi{10.1093/mnras/stu2089}

\bibitem[{{Me{\v{s}}tri{\'c}} {et~al.}(2021){Me{\v{s}}tri{\'c}}, {Ryan-Weber},
  {Cooke}, {Bassett}, {Prichard}, \& {Rafelski}}]{Mestric2021}
{Me{\v{s}}tri{\'c}}, U., {Ryan-Weber}, E.~V., {Cooke}, J., {et~al.} 2021,
  \mnras, 508, 4443, \dodoi{10.1093/mnras/stab2615}

\bibitem[{{Mitra} {et~al.}(2015){Mitra}, {Choudhury}, \& {Ferrara}}]{Mitra15}
{Mitra}, S., {Choudhury}, T.~R., \& {Ferrara}, A. 2015, \mnras, 454, L76,
  \dodoi{10.1093/mnrasl/slv134}

\bibitem[{{Naidu} {et~al.}(2022){Naidu}, {Matthee}, {Oesch}, {Conroy},
  {Sobral}, {Pezzulli}, {Hayes}, {Erb}, {Amor{\'\i}n}, {Gronke}, {Schaerer},
  {Tacchella}, {Kerutt}, {Paulino-Afonso}, {Calhau}, {Llerena}, \&
  {R{\"o}ttgering}}]{Naidu2022}
{Naidu}, R.~P., {Matthee}, J., {Oesch}, P.~A., {et~al.} 2022, \mnras, 510,
  4582, \dodoi{10.1093/mnras/stab3601}

\bibitem[{{Nelson} {et~al.}(2015){Nelson}, {Pillepich}, {Genel},
  {Vogelsberger}, {Springel}, {Torrey}, {Rodriguez-Gomez}, {Sijacki}, {Snyder},
  {Griffen}, {Marinacci}, {Blecha}, {Sales}, {Xu}, \& {Hernquist}}]{Nelson2015}
{Nelson}, D., {Pillepich}, A., {Genel}, S., {et~al.} 2015, Astronomy and
  Computing, 13, 12, \dodoi{10.1016/j.ascom.2015.09.003}

\bibitem[{{Norris} {et~al.}(2010){Norris}, {Wyse}, {Gilmore}, {Yong}, {Frebel},
  {Wilkinson}, {Belokurov}, \& {Zucker}}]{Norris2010}
{Norris}, J.~E., {Wyse}, R. F.~G., {Gilmore}, G., {et~al.} 2010, \apj, 723,
  1632, \dodoi{10.1088/0004-637X/723/2/1632}

\bibitem[{{O{\~n}orbe} {et~al.}(2017){O{\~n}orbe}, {Hennawi}, \&
  {Luki{\'c}}}]{Onorbe2017}
{O{\~n}orbe}, J., {Hennawi}, J.~F., \& {Luki{\'c}}, Z. 2017, \apj, 837, 106,
  \dodoi{10.3847/1538-4357/aa6031}

\bibitem[{{Omukai}(2000)}]{Omukai2000}
{Omukai}, K. 2000, ApJ, 534, 809, \dodoi{10.1086/308776}

\bibitem[{{Osterbrock} \& {Ferland}(2006)}]{Osterbrock2006}
{Osterbrock}, D.~E., \& {Ferland}, G.~J. 2006, {Astrophysics of gaseous nebulae
  and active galactic nuclei (CA: University Science Books)}, ed. {Osterbrock,
  D.~E.~\& Ferland, G.~J.}

\bibitem[{{Padmanabhan}(1993)}]{Padmanabhan1993}
{Padmanabhan}, T. 1993, {Structure Formation in the Universe}

\bibitem[{{Patel} {et~al.}(2017){Patel}, {Besla}, \& {Sohn}}]{Patel2017}
{Patel}, E., {Besla}, G., \& {Sohn}, S.~T. 2017, \mnras, 464, 3825,
  \dodoi{10.1093/mnras/stw2616}

\bibitem[{{Patel} {et~al.}(2020){Patel}, {Kallivayalil}, {Garavito-Camargo},
  {Besla}, {Weisz}, {van der Marel}, {Boylan-Kolchin}, {Pawlowski}, \&
  {G{\'o}mez}}]{Patel2020}
{Patel}, E., {Kallivayalil}, N., {Garavito-Camargo}, N., {et~al.} 2020, \apj,
  893, 121, \dodoi{10.3847/1538-4357/ab7b75}

\bibitem[{{Pawlik} \& {Schaye}(2008)}]{Pawlik2008}
{Pawlik}, A.~H., \& {Schaye}, J. 2008, MNRAS, 389, 651,
  \dodoi{10.1111/j.1365-2966.2008.13601.x}

\bibitem[{{Pereira-Wilson} {et~al.}(2023){Pereira-Wilson}, {Navarro},
  {Ben{\'\i}tez-Llambay}, \& {Santos-Santos}}]{Pereira23}
{Pereira-Wilson}, M., {Navarro}, J.~F., {Ben{\'\i}tez-Llambay}, A., \&
  {Santos-Santos}, I. 2023, \mnras, 519, 1425, \dodoi{10.1093/mnras/stac3633}

\bibitem[{{Planck Collaboration}(2016)}]{planck2016}
{Planck Collaboration}. 2016, A\&A, 594, A13,
  \dodoi{10.1051/0004-6361/201525830}

\bibitem[{{Portinari} {et~al.}(1998){Portinari}, {Chiosi}, \&
  {Bressan}}]{Portinari1998}
{Portinari}, L., {Chiosi}, C., \& {Bressan}, A. 1998, A\&A, 334, 505

\bibitem[{{Puchwein} {et~al.}(2015){Puchwein}, {Bolton}, {Haehnelt}, {Madau},
  {Becker}, \& {Haardt}}]{Puchwein2015}
{Puchwein}, E., {Bolton}, J.~S., {Haehnelt}, M.~G., {et~al.} 2015, \mnras, 450,
  4081, \dodoi{10.1093/mnras/stv773}

\bibitem[{{Rey} {et~al.}(2020){Rey}, {Pontzen}, {Agertz}, {Orkney}, {Read}, \&
  {Rosdahl}}]{Rey2020}
{Rey}, M.~P., {Pontzen}, A., {Agertz}, O., {et~al.} 2020, arXiv:2004.09530,
  arXiv:2004.09530.
\newblock \doarXiv{2004.09530}

\bibitem[{{Rosdahl} {et~al.}(2018){Rosdahl}, {Katz}, {Blaizot}, {Kimm},
  {Michel-Dansac}, {Garel}, {Haehnelt}, {Ocvirk}, \& {Teyssier}}]{Rosdahl18}
{Rosdahl}, J., {Katz}, H., {Blaizot}, J., {et~al.} 2018, \mnras, 479, 994,
  \dodoi{10.1093/mnras/sty1655}

\bibitem[{{Sacchi} {et~al.}(2021){Sacchi}, {Richstein}, {Kallivayalil}, {van
  der Marel}, {Libralato}, {Zivick}, {Besla}, {Brown}, {Choi}, {Deason},
  {Fritz}, {Geha}, {Guhathakurta}, {Jeon}, {Kirby}, {Majewski}, {Patel},
  {Simon}, {Tony Sohn}, {Tollerud}, \& {Wetzel}}]{Sachhi2021}
{Sacchi}, E., {Richstein}, H., {Kallivayalil}, N., {et~al.} 2021, \apjl, 920,
  L19, \dodoi{10.3847/2041-8213/ac2aa3}

\bibitem[{{Safranek-Shrader} {et~al.}(2016){Safranek-Shrader}, {Montgomery},
  {Milosavljevi{\'c}}, \& {Bromm}}]{Safranek2016}
{Safranek-Shrader}, C., {Montgomery}, M.~H., {Milosavljevi{\'c}}, M., \&
  {Bromm}, V. 2016, \mnras, 455, 3288, \dodoi{10.1093/mnras/stv2545}

\bibitem[{{Sanati} {et~al.}(2023){Sanati}, {Jeanquartier}, {Revaz}, \&
  {Jablonka}}]{Sanati2023}
{Sanati}, M., {Jeanquartier}, F., {Revaz}, Y., \& {Jablonka}, P. 2023, \aap,
  669, A94, \dodoi{10.1051/0004-6361/202244309}

\bibitem[{{Santistevan} {et~al.}(2020){Santistevan}, {Wetzel}, {El-Badry},
  {Bland-Hawthorn}, {Boylan-Kolchin}, {Bailin}, {Faucher-Gigu{\`e}re}, \&
  {Benincasa}}]{Santistevan2020}
{Santistevan}, I.~B., {Wetzel}, A., {El-Badry}, K., {et~al.} 2020, \mnras, 497,
  747, \dodoi{10.1093/mnras/staa1923}

\bibitem[{{Sawala} {et~al.}(2010){Sawala}, {Scannapieco}, {Maio}, \&
  {White}}]{Sawala2011}
{Sawala}, T., {Scannapieco}, C., {Maio}, U., \& {White}, S. 2010, MNRAS, 402,
  1599, \dodoi{10.1111/j.1365-2966.2009.16035.x}

\bibitem[{{Schmidt}(1959)}]{Schmidt1959}
{Schmidt}, M. 1959, ApJ, 129, 243, \dodoi{10.1086/146614}

\bibitem[{{Schneider} \& {Omukai}(2010)}]{Schneider2010}
{Schneider}, R., \& {Omukai}, K. 2010, MNRAS, 402, 429,
  \dodoi{10.1111/j.1365-2966.2009.15891.x}

\bibitem[{{Simon}(2019)}]{Simon2019}
{Simon}, J.~D. 2019, ARA\&A, 57, 375,
  \dodoi{10.1146/annurev-astro-091918-104453}

\bibitem[{{Simpson} {et~al.}(2013){Simpson}, {Bryan}, {Johnston}, {Smith}, {Mac
  Low}, {Sharma}, \& {Tumlinson}}]{Simpson2013}
{Simpson}, C.~M., {Bryan}, G.~L., {Johnston}, K.~V., {et~al.} 2013, \mnras,
  432, 1989, \dodoi{10.1093/mnras/stt474}

\bibitem[{{Springel}(2005)}]{Springel2005}
{Springel}, V. 2005, MNRAS, 364, 1105, \dodoi{10.1111/j.1365-2966.2005.09655.x}

\bibitem[{{Springel} {et~al.}(2001){Springel}, {White}, {Tormen}, \&
  {Kauffmann}}]{Springel2001}
{Springel}, V., {White}, S.~D.~M., {Tormen}, G., \& {Kauffmann}, G. 2001,
  MNRAS, 328, 726, \dodoi{10.1046/j.1365-8711.2001.04912.x}

\bibitem[{{Stinson} {et~al.}(2007){Stinson}, {Dalcanton}, {Quinn}, {Kaufmann},
  \& {Wadsley}}]{Stinson2007}
{Stinson}, G.~S., {Dalcanton}, J.~J., {Quinn}, T., {Kaufmann}, T., \&
  {Wadsley}, J. 2007, ApJ, 667, 170, \dodoi{10.1086/520504}

\bibitem[{{Suda} {et~al.}(2008){Suda}, {Katsuta}, {Yamada}, {Suwa}, {Ishizuka},
  {Komiya}, {Sorai}, {Aikawa}, \& {Fujimoto}}]{Suda2008}
{Suda}, T., {Katsuta}, Y., {Yamada}, S., {et~al.} 2008, \pasj, 60, 1159,
  \dodoi{10.1093/pasj/60.5.1159}

\bibitem[{{Tolstoy} {et~al.}(2009){Tolstoy}, {Hill}, \& {Tosi}}]{Tolstoy2009}
{Tolstoy}, E., {Hill}, V., \& {Tosi}, M. 2009, ARA\&A, 47, 371,
  \dodoi{10.1146/annurev-astro-082708-101650}

\bibitem[{{Vogelsberger} {et~al.}(2014){Vogelsberger}, {Genel}, {Springel},
  {Torrey}, {Sijacki}, {Xu}, {Snyder}, {Nelson}, \&
  {Hernquist}}]{Vogelsberger2014}
{Vogelsberger}, M., {Genel}, S., {Springel}, V., {et~al.} 2014, \mnras, 444,
  1518, \dodoi{10.1093/mnras/stu1536}

\bibitem[{{Weisz} {et~al.}(2014){Weisz}, {Dolphin}, {Skillman}, {Holtzman},
  {Gilbert}, {Dalcanton}, \& {Williams}}]{Weisz2014}
{Weisz}, D.~R., {Dolphin}, A.~E., {Skillman}, E.~D., {et~al.} 2014, \apj, 789,
  147, \dodoi{10.1088/0004-637X/789/2/147}

\bibitem[{{Weisz} {et~al.}(2019){Weisz}, {Martin}, {Dolphin}, {Albers},
  {Collins}, {Ferguson}, {Lewis}, {Mackey}, {McConnachie}, {Rich}, \&
  {Skillman}}]{Weisz2019}
{Weisz}, D.~R., {Martin}, N.~F., {Dolphin}, A.~E., {et~al.} 2019, \apjl, 885,
  L8, \dodoi{10.3847/2041-8213/ab4b52}

\bibitem[{{Wheeler} {et~al.}(2019){Wheeler}, {Hopkins}, {Pace},
  {Garrison-Kimmel}, {Boylan-Kolchin}, {Wetzel}, {Bullock}, {Kere{\v{s}}},
  {Faucher-Gigu{\`e}re}, \& {Quataert}}]{Wheeler2019}
{Wheeler}, C., {Hopkins}, P.~F., {Pace}, A.~B., {et~al.} 2019, MNRAS, 490,
  4447, \dodoi{10.1093/mnras/stz2887}

\bibitem[{{Wiersma} {et~al.}(2009){Wiersma}, {Schaye}, {Theuns}, {Dalla
  Vecchia}, \& {Tornatore}}]{Wiersma2009b}
{Wiersma}, R.~P.~C., {Schaye}, J., {Theuns}, T., {Dalla Vecchia}, C., \&
  {Tornatore}, L. 2009, MNRAS, 399, 574,
  \dodoi{10.1111/j.1365-2966.2009.15331.x}

\bibitem[{{Wyithe} \& {Loeb}(2003)}]{Wyite2003}
{Wyithe}, J. S.~B., \& {Loeb}, A. 2003, \apj, 586, 693, \dodoi{10.1086/367721}

\bibitem[{{Yajima} {et~al.}(2011){Yajima}, {Choi}, \& {Nagamine}}]{Yajima11}
{Yajima}, H., {Choi}, J.-H., \& {Nagamine}, K. 2011, \mnras, 412, 411,
  \dodoi{10.1111/j.1365-2966.2010.17920.x}

\end{thebibliography}
\bibliographystyle{aasjournal}



\end{document}